\documentclass[aps,pre,preprint,showpacs,showkeys]{revtex4}
\usepackage{amssymb}
\usepackage{amsfonts}
\usepackage{makeidx}
\usepackage{graphicx}
\usepackage{amsmath}
\usepackage{appendix}

\begin{document}

\author{K. P. Santo}
\email{poulose@ualberta.ca}
\altaffiliation{currently at National Institute of Nano-technology, NRC, Canada}
\author{K. L. Sebastian}
\email{kls@ipc.iisc.ernet.in}
\title{\textbf{The dynamics of loop formation in a semiflexible polymer }}
\date{\today }

\begin{abstract}
The dynamics of loop formation by linear polymer chains has been a topic of
several theoretical/experimental studies. Formation of loops and their
opening are key processes in many important biological processes. Loop
formation in flexible chains has been extensively studied by many groups.
However, in the more realistic case of semiflexible polymers, not much
results are available. In a recent study (K. P. Santo and K. L. Sebastian,
Phys. Rev. E, \textbf{73}, 031293 (2006)), we investigated opening dynamics
of semiflexible loops in the short chain limit and presented results for
opening rates as a function of the length of the chain. We presented an
approximate model for a semiflexible polymer in the rod limit, based on a
semiclassical expansion of the bending energy of the chain. The model
provided an easy way to describe the dynamics. In this paper, using this
model, we investigate the reverse process, i.e., the loop formation dynamics
of a semiflexible polymer chain by describing the process as a
diffusion-controlled reaction. We make use of the \textquotedblleft closure
approximation" of Wilemski and Fixmann (G. Wilemski and M. Fixmann, J. Chem.
Phys., \textbf{60}, 878 (1974)), in which a sink function is used to
represent the reaction. We perform a detailed multidimensional analysis of
the problem and calculate closing times for a semiflexible chain. We show
that for short chains, the loop formation time $\tau $ decreases with the
contour length of the polymer. But for longer chains, it increases with
length obeying a power law and so it has a minimum at an intermediate
length. \ In terms of dimensionless variables, the closing time is found to
be given by $\tau \sim L^{n}\exp (\mathrm{const.}/L)$, where $n=4.5-6$. The
minimum loop formation time occurs at a length \ $L_{m}$ of about $2.2-2.4$.
These are, indeed, the results that are physically expected, but a
multidimensional analysis leading to these results does not seem to exist in
the literature so far.
\end{abstract}

\affiliation{Department of Inorganic and Physical chemistry, Indian Institute of
Science,
Bangalore, India}
\maketitle




\section{Introduction}

In a previous paper \cite{klssantoPRE2006}, the dynamics of opening of a
weak bond between the two ends of a semiflexible polymer chain was
considered in detail. An approximate model for a stiff polymer ring in the
rod limit, based on a \textquotedblleft semiclassical" method was developed.
This model, though approximate, was found to provide an easy approach to
describe the dynamics of a worm-like polymer chain in the rod limit. In Ref.
\cite{klssantoPRE2006}, we used the model to analyze the dynamics of opening
and to calculate the rates of opening as a function of length in the short
chain limit. Here in this paper, we analyze the dynamics of loop formation.

The closing dynamics of polymer chains has been studied extensively, being
the key process in important biological functions, such as control of gene
expression \cite{rippe2001,rippe1995}, DNA replication \cite{junPC2003} and
protein folding. Experimental studies on loop formation involve monitoring
the dynamics of DNA hairpins \cite%
{libchaber98PNAS8602,wallace01PNAS8602,Libchaber00PRL2400,AnsariPNAS2001,
AnsariJPCB2001} and small peptides \cite%
{szaboJPCB2002,szaboPNAS1998,LapidusPNAS2000,LapidusPRL2001}, using
fluorescence spectroscopic techniques. Several theoretical approaches are
available for analyzing loop formation of a flexible chain. Using the
formalism of Wilemski and Fixman (WL) \cite{wilemski} for diffusion
controlled reactions, the closing time $\tau $ for a flexible chain was
calculated by Doi \cite{Doi75CP455} and was found to vary as, $\tau \sim
L^{2}$. In another important approach, Szabo, Schulten and Schulten (SSS)
\cite{shultenJCP1980} calculated the mean first passage time for closing for
a gaussian chain and found $\tau \sim L^{3/2}$. The two approaches have been
analyzed by recent simulations \cite{srinivasJCP2002,pastor96JCP3878}. But
real polymers such as DNA, RNA and proteins are not flexible and hence, it
is more important to understand the closing dynamics of stiff chains.
Unfortunately, in this case, only simple, approximate approaches \cite%
{freyPRL2003,wangPRE2000,winkler94JCP8119,freedJCP1971,harisJCp1966,
saitoJPSC1967}
are available in the literature so far. Since worm-like chains are
represented by differentiable curves, one has to incorporate the constraint $%
|\mathbf{u}(s)|=1$ and this has been a problem in dealing with semiflexible
polymers. Yamakawa and Stockmayer \cite{stockmayer72JCP2843} and Shimada and
Yamakawa \cite{shimadaJCP1984} have calculated the static ring closure
probabilities for worm-like chains and helical worm-like chains. According
to their analysis, the ring closure probability for a worm-like chain has
the form, $G(\mathbf{0};L)=896.32(l_{p}/L)^{5}\exp
(-14.054l_{p}/L+0.246L/l_{p}),$ where $l_{p}$ is the persistence length of
the chain. An approximate treatment that leads to the end-to-end probability
distribution for semiflexible polymers has been given by Winkler et. al \cite%
{winkler94JCP8119} and using their approach, the closing dynamics has been
analyzed recently by Cherayil and Dua \cite{cherayil02JCP7765}. They find
that the closing time $\tau \sim L^{\nu }$, where $\nu $ is in the range $%
2.2 $ to $2.4$. In an interesting paper, Jun. \textit{et. al} \cite%
{junEPL2003} showed that the closing time should decrease with length in the
short chain limit and then increase with length for longer chains. Hence,
the closing time has a minimum at an intermediate length. The reason for
this behavior is that, for short chains, the bending energy contributes
significantly to the activation energy for the process. Thus the activation
energy $\sim {\mathrm{const.}}/L$ and therefore the closing time $\tau \sim
\exp ({\mathrm{const.}}/L)$. For longer chains, the free energy barrier for
closing is due to the configurational entropy and hence, $\tau $ obeys a
power law. Jun \textit{et. al} \cite{junEPL2003} have followed an
approximate one dimensional Kramers approach to reproduce this behavior and
obtain the minimum of closing time at a length $L_{m}=3.4l_{p}$, where $%
l_{p} $ is the persistence length of the chain. Monte Carlo simulations by
Chen \textit{et.al} \cite{chen04EL407} lead to $L_{m}=2.85l_{p}$. See also
the paper by Ranjith et al \cite{SunilKumar2005}.

In Ref. \cite{klssantoPRE2006}, we analyzed the opening dynamics of a
semiflexible polymer ring formed by a weak bond between the ends. We
developed a model that describe the polymer near the ring configuration,
using a semiclassical expansion of the bending energy of the chain. The
model, though approximate, provided an easy way to analyze the dynamics.
Using this model, we calculated the opening rates as function of the contour
length of the chain. The formalism presented in Ref.\cite{klssantoPRE2006},
took into account of the inextensibility constraint, $|\mathbf{u}(s)|=1$ for
semiflexible chains rigourously. The conformations of the chain can be
mapped onto the paths of a Brownian particle on a unit sphere. We performed
a semiclassical expansion about the most probable path assuming that the
fluctuations about the most probable path are small. For the ring, we took
most probable path to be the great circle on the sphere. This is again an
approximation, as the minimum energy configuration for a semiflexible
polymer loop does not correspond to the great circle. However, as described
in Ref. \cite{klssantoPRE2006}, it led to minimum energy values very close
to exact results by Yamakawa and Stockmayer \cite{stockmayer72JCP2843} and
the approximation scheme by Kulic and Schiessel \cite{kulic2003}. Once the
ends of a semiflexible polymer are brought together, they can separate in
any of the three directions in space. Our analysis showed that two of the
three directions in space are unstable, while one direction is stable. If
one considers the ring to be in the $XY$-plane, with its ends meeting on the
$Y$ axis, then the motion that leads to separation along the $Y$ direction
is stable, while the motions lead to separation along $X$ or $Z$ direction
are unstable. The nature of instabilities along the $X$ and $Z$ directions
are different. Hence, near the ring, the three directions in space are
non-equivalent for a semiflexible polymer and are governed by different
energetics (see Sec. \ref{sectwo2}). One may also perform the expansion near
the rod configuration by expanding about the straight rod. On the unit
sphere the straight rod corresponds to a point and unlike the great circle
this is an exact minimum energy configuration (see Sec. \ref{semiclassical}).

In this paper, we present a detailed multidimensional analysis of the
dynamics of loop formation in semiflexible chains. We make use of the
approximation scheme developed in Ref. \cite{klssantoPRE2006}. Following
Wilemski and Fixman \cite{wilemski}, the looping is described as a
diffusion-controlled reaction. In the WF theory, the effect of the reaction
is incorporated into the model using a sink function. In special cases,
exact analytical results are possible for a delta function sink
\cite{bicoutJCP1997,klsPRA1992,Ananya2006}. But for an arbitrary
sink, and multidimensional dynamics, this is not possible. For such cases,
WF suggested an approximation known as the \textquotedblleft closure"
approximation. In this, the diffusion limited life-time of the process is
expressed in terms of a sink-sink correlation function and the essential
step for finding the loop formation time is to calculate this sink-sink
correlation function. For this, we need to know the time-dependent Green's
function of the chain and the equilibrium probability distribution. We
therefore derive the time-dependent multidimensional Green's function of the
semiflexible polymer near the loop configuration by performing a normal mode
analysis. This Green's function is then used to find the sink-sink
correlation function for a Gaussian sink and the closing time. We find that
the closing time $\tau \sim (L/l_{p})^{n}\exp (Al_{p}/L)$, The exponent $%
n=4.5$ - $6$. $\tau $ is found to be a minimum at a length $L_{m}\simeq $ $%
2.2-2.4l_{p}$ which has to be compared with the value $3.4l_{p}$ obtained in
Ref. \cite{junEPL2003} and $2.85l_{p}$ of Ref. \cite{chen04EL407}. We find $%
L_{m}$ to be weakly dependent on the range of the interaction between the
ends. Thus, our analysis leads to results that are physically expected. It
is worth mentioning that a multidimensional analysis leading to these
results does not seem to exist in the literature so far. \ We also calculate
the loop formation probability $G(\mathbf{0};L)$ and find that our method
leads to the correct behavior, i.e., $G(\mathbf{0};L)\sim L^{-5}\exp (-%
\mathrm{const.}/L)$, thus showing that the procedure reproduces the previous
results for this quantity \cite{shimadaJCP1984}.

The paper is organized as follows: In Sec.\ref{Closure}, we give a summary
of the WF theory for diffusion controlled reactions and the
\textquotedblleft closure" approximation. In Sec. \ref{semiclassical}, the
semiclassical approximation scheme for bending energy of a semiflexible
polymer is briefly outlined. The time dependent Green's function of the
polymer is derived though a normal mode analysis near the loop configuration
in Sec. \ref{NormalCoordinates}. The approximate probability distribution of
the chain is given in Sec. \ref{NormalCoordinates}. In Sec. \ref%
{sinkcorrelationfunction}, we calculate the sink-sink correlation function
for a Gaussian sink and the closing time. In Sec. \ref{numericalresults}, we
give numerical results. Summary and conclusions are given in Sec. \ref%
{secsix}.

\bigskip

\section{The ``closure" approximation}

\label{Closure} In this section, we summarize the theory of
diffusion-controlled intra-chain reactions of polymers developed by Wilemski
and Fixman \cite{wilemski} and their \textquotedblleft\ closure"
approximation for an arbitrary sink function. The dynamics of a single
polymer chain in a viscous environment is governed by the diffusion
equation,
\begin{equation}
\frac{\partial P}{\partial t}+{\hat{D}}P=0,  \label{diffeq}
\end{equation}%
where ${\hat{D}}$ is the diffusion operator for the chain. If the chain is
represented by $N+1$ beads with position vectors represented by $\mathbf{r}=(%
\mathbf{r}_{1},\mathbf{r}_{2},....\mathbf{r}_{N+1})$, then the general form
of the diffusion operator is given by
\begin{equation}
{\hat{D}}=D\sum\limits_{i=1}^{N+1}\nabla _{i}.(\nabla _{i}+(k_{B}T)^{-1}%
\mathbf{F}_{i}).  \label{difop}
\end{equation}%
$D=k_{B}T/\xi $, is the diffusion coefficient of the segments and $\xi $ is
the friction coefficient of the segments. $\mathbf{F}_{i}=\nabla _{i}U$,
where $U$ is the potential energy of the chain. Eq. (\ref{diffeq}) may be
solved to obtain the equilibrium distribution $P_{eq}$ of the chain, which
is time independent. But if the chain has reactive ends, they can react and
form a loop when they come sufficiently close and hence the probability
distribution of an open chain will decay in time. In such a case, one may
solve Eq. (\ref{diffeq}) with appropriate boundary conditions. An alternate
approach to the same problem is to introduce a sink function into the
equation for $P(\mathbf{r},t)$ as done by Wilemski and Fixman \cite{wilemski}%
. Then the reaction-diffusion equation that governs the dynamics of a
polymer chain with reactive ends is
\begin{equation}
\frac{\partial P}{\partial t}+{\hat{D}}P=-k_{r}\mathcal{S}(\mathbf{r})P,
\label{difsink}
\end{equation}%
where $P$ is the distribution function of the open polymer chain and $k_{r}$
is the strength of the sink function. $k_{r}$ determines the rate at which
the reaction occurs when the ends are sufficiently close. $\mathcal{S}$ is
the sink function and is a function of $\mathbf{r}_{1},\mathbf{r}_{2},...%
\mathbf{r}_{N+1}$. Integrating Eq. (\ref{difsink}) over all the coordinates $%
\mathbf{r}$ we get
\begin{equation}
\frac{dP_{s}(t)}{dt}=-k_{r}v(t),  \label{surveq}
\end{equation}%
where
\begin{equation}
v(t)=\int d\mathbf{r}\mathcal{S}(\mathbf{r})P(\mathbf{r},t)  \label{vt}
\end{equation}%
and
\begin{equation}
P_{s}(t)=\int d\mathbf{r}P(\mathbf{r},t)  \label{surprob}
\end{equation}%
is the survival probability. The function $\mathcal{S}$ can be any suitable
function, but is usually taken to be a delta function or a gaussian. Eq. (%
\ref{difsink}) can be solved exactly only in one dimension for a delta
function sink or a quadratic sink (see references
\cite{bicoutJCP1997,klsPRA1992,Ananya2006}
, and the references therein). Therefore, WF introduced the assumption that $%
P(\mathbf{r},t)$ may be approximated as
\begin{equation}
P(\mathbf{r},t)=P_{eq}(\mathbf{r})\nu (t),  \label{approx}
\end{equation}%
where
\begin{equation}
\nu (t)=\frac{v(t)}{v_{eq}}  \label{nut}
\end{equation}%
with
\begin{equation}
v_{eq}=\int d\mathbf{r}\mathcal{S}(\mathbf{r})P_{eq}(\mathbf{r}).
\label{veq}
\end{equation}%
This is referred to as the \textquotedblleft closure" approximation. The
average time of closing is the integral of the survival probability and is
given by
\begin{equation}
\tau =\int\limits_{0}^{\infty }P_{s}(t)dt=\tilde{P_{s}}(0),  \label{tau}
\end{equation}%
where $\tilde{P_{s}}(s)$ is the Laplace transform of $P_{s}(t)$. $\tau $ is
also expressed in terms of a sink-sink correlation function and in the
diffusion-limited ($k_{r}\rightarrow \infty $) limit, it is given by \cite%
{wilemski}
\begin{equation}
\tau =\int\limits_{0}^{\infty }\left( \frac{\mathcal{D}(t)}{v_{eq}^{2}}%
-1\right) dt.  \label{sinktau}
\end{equation}%
$\mathcal{D}(t)$ is the sink-sink correlation function defined by
\begin{equation}
\mathcal{D}(t)=\int d\mathbf{r}\int d\mathbf{r}^{\prime }\mathcal{S}(\mathbf{%
r})G_{0}(\mathbf{r},\mathbf{r}^{\prime };t)\mathcal{S}(\mathbf{r}^{\prime
})P_{eq}(\mathbf{r}^{\prime }),  \label{sinkcor}
\end{equation}%
where $G_{0}(\mathbf{r},\mathbf{r}^{\prime };t)$ is the Green's function for
the diffusive motion of the chain in the absence of the sink. Eq. (\ref%
{sinktau}) was obtained by WF \cite{wilemski}. Note that $\mathcal{D}(\infty
)=v_{eq}^{2}$. To calculate $\mathcal{D}(t)$, one needs to know $G_{0}(%
\mathbf{r},\mathbf{r}^{\prime };t)$ and $P_{eq}(\mathbf{r})$ and these will
be calculated in the following sections. We shall take $\mathcal{S}(\mathbf{r%
})$ to be a Gaussian, given by%
\begin{equation}
\mathcal{S}(\mathbf{r})\equiv \mathcal{S}(\mathbf{R})=\mathcal{S}_{x}(R_{x})%
\mathcal{S}_{y}(R_{y})\mathcal{S}_{z}(R_{z}),  \label{S(R)}
\end{equation}%
where $\mathbf{R}$ is the end to end vector for the chain and

\begin{equation}
\mathcal{S}_{i}(R_{i})=e^{-R_{i}^{2}/(2\eta ^{2})}/(\sqrt{2\pi }\ \eta ),
\text{ \ }i=x,y\text{ \ or }z .  \label{SiRi}
\end{equation}%
$\eta $ is the width of the Gaussian sink.

\section{The semi-classical Approximation scheme for the Bending Energy}

\label{semiclassical} 
\begin{figure}[tbp]
\includegraphics[height=0.5\linewidth,keepaspectratio=true]{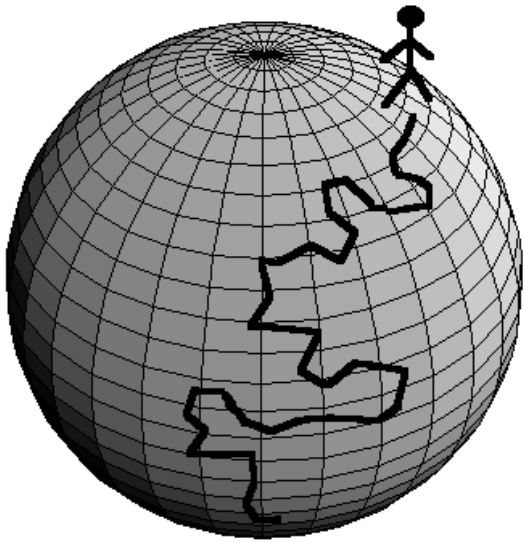}
\caption{The conformations of a semiflexible polymer may be mapped onto the
paths of a Brownian particle on the surface of a unit sphere. The circular
ring polymer with the tangent vectors at the ends joining smoothly
corresponds to the great circle on the unit sphere.}
\label{brsphere}
\end{figure}

In Ref.\cite{klssantoPRE2006}, we introduced an
approximation scheme for the bending energy of a semiflexible polymer ring,
which is based on a ``semiclassical" expansion. In this section, we give a
brief account of the approach. A semiflexible polymer is usually considered
as a continuous, inextensible space curve represented by the position vector
$\mathbf{r}(s)$, where $s$ is the arc-length parameter. The bending energy
of the chain is given by
\begin{equation}  \label{Ebend}
E_{bend}=\frac{\kappa}{2}\int\limits_{0}^L \left(\frac{\partial^2 \mathbf{r}%
(s)}{\partial s^2}\right)^2ds.
\end{equation}
$\kappa$ is the bending rigidity. Since the curve is differentiable, one has
the constraint,
\begin{equation}  \label{cons}
|\mathbf{u}(s)|=1,
\end{equation}
where $\mathbf{u}(s)=\partial \mathbf{r}(s)/\partial s$, the tangent vector
at the point $s$. The partition function of the semiflexible polymer is the
functional integral over the conformations represented by $\mathbf{r}(s)$,
\begin{equation}  \label{partfn}
Z=\int \mathbf{Dr}(s)\exp\left(\frac{-E_{bend}[\mathbf{r}(s)]}{k_B T}\right).
\end{equation}
This functional integral has to be performed with the constraint of Eq. (\ref%
{cons}). However, incorporating this constraint has been a problem in
dealing with semiflexible polymers. In Ref. \cite{klssantoPRE2006}, we wrote
the partition function as an integral over ${\mathbf{u}}(s)$,
\begin{equation}  \label{pfu}
Z=\int \mathbf{Du}(s)\exp\left(\frac{-E_{bend}[\mathbf{u}(s)]}{k_B T}\right).
\end{equation}
and represented $\mathbf{u}(s)$ in angle coordinates
\begin{equation}
\mathbf{u}(s)=\mathbf{i}\sin \theta (s)\cos \phi (s)+\mathbf{j}\sin \theta
(s)\sin \phi (s)+\mathbf{k}\cos \theta (s).  \label{tangent}
\end{equation}
Since the magnitude of the tangent vector is one, the conformations of the
semiflexible polymer can be mapped onto the trajectories of a Brownian
particle over a unit sphere (Fig.\ref{brsphere}).
The bending energy of the chain is then written in terms of the angles $%
\theta$ and $\phi$ as
\begin{equation}
E_{bend}=\frac{\kappa }{2}\int\limits_{0}^{L}ds\left\{ \left( \frac{d\theta
(s)}{ds}\right) ^{2}+\sin ^{2}\theta (s)\left( \frac{d\phi (s)}{ds}\right)
^{2}\right\}  \label{Ebend1}
\end{equation}
and the partition function is written as a path integral in spherical polar
coordinates
\begin{equation}
Z_{loop}=\int D\theta (s)\int D\phi (s)\exp \left( -\frac{ E_{bend}[\theta
(s),\phi (s)]}{k_B T}\right).  \label{pfthetaphi}
\end{equation}
This path integral has not been evaluated in a closed form. In Ref. \cite%
{klssantoPRE2006}, we have used a semiclassical expansion of the bending
energy, to evaluate the above partition function approximately.

\subsection{Bending energy of the loop: expansion about the great circle}

\label{BendingenergyLoop} To perform a semiclassical expansion of the
bending energy of the semiflexible polymer near the ring configuration, we
take the most important path to be the great circle on the unit sphere (Fig. %
\ref{brsphere}). The great circle corresponds to a ring with the tangents
smoothly joined. However, the minimum energy configuration of a rod-like
polymer whose ends are brought together to form a loop would not have its
tangents joining smoothly and therefore does not correspond to a great
circle \cite{stockmayer72JCP2843}. Hence, our approach is approximate but
has the advantage that it provides an easy way to study the dynamics. \ On
the other hand, if one is interested in covalent bond formation, in which
directionality of the bond is important, then the great circle is the
appropriate starting point.

The position vector of the polymer may be found by inverting the definition
of the tangent vector ${\mathbf{u}}(s)=\partial {\mathbf{r}}(s)/\partial s$,
\begin{equation}
\mathbf{r}(s)=\mathbf{r}_{cm}-\frac{1}{L}\int_{0}^{L}ds\int_{0}^{s}ds_{1}%
\mathbf{u}(s_{1})+\int_{0}^{s}ds_{1}\mathbf{u}(s_{1}),
\label{positionvector}
\end{equation}%
where $\mathbf{r}_{cm}$ denotes the position vector of the center of mass of
the ring polymer. The great circle is chosen to lie in the XY plane of a
Cartesian coordinate system, with any point on it represented by the
coordinates, $[\theta (s),\phi (s)]=\left[ \pi /2,2\pi s/L\right] $. The
position vector of the circular ring polymer that corresponds to the great
circle may be found using Eq. (\ref{positionvector}) and is given by
\begin{equation}
\mathbf{r}_{GC}(s)=\frac{L}{2\pi }\left[ \mathbf{i}\sin \left( \frac{2\pi s}{%
L}\right) -\mathbf{j}\cos \left( \frac{2\pi s}{L}\right) \right] .
\label{rexm}
\end{equation}%
This curve represents one end of the polymer lying in the XY-plane starting
at $\frac{-L}{2\pi }$ on the negative Y axis, going around the Z axis along
a circle of radius $\frac{L}{2\pi }$, coming back to the same point after
traversing a circle of radius $\frac{L}{2\pi }$. The fluctuations about this
path are taken into account by letting
\begin{equation}
\lbrack \theta (s),\phi (s)]=\left[ \frac{\pi }{2}+\delta \theta (s),\frac{%
2\pi s}{L}+\delta \phi (s)\right] ,  \label{fluctuations}
\end{equation}%
where $\delta \theta (s)$ and $\delta \phi (s)$ represent the deviations
from the extremum path on the unit sphere expressed in terms of angles.
Expanding the bending energy of Eq. (\ref{Ebend1}) correct up to second
order in the fluctuations $\delta \theta (s)$ and $\delta \phi (s)$ gives
\begin{equation}
\begin{array}{l}
E_{bend}=\frac{\kappa }{2}\int\limits_{0}^{L}ds\left\{ \left( \frac{d\delta
\theta (s)}{ds}\right) ^{2}+\left( \frac{2\pi }{L}+\frac{d\delta \phi (s)}{ds%
}\right) ^{2}-\left( \frac{2\pi }{L}\right) ^{2}\delta \theta
^{2}(s)\right\} .\label{secordE}%
\end{array}%
\end{equation}%
We expect this expansion to be a valid approximation near the ring
configuration, if the deviations from the circular configuration is small.

We expand fluctuations as
\begin{equation}
\delta \phi (s)=\sum\limits_{n=0}^{\infty }\delta \phi _{n}\cos \left( \frac{%
n\pi s}{L}\right)  \label{deltaphi}
\end{equation}%
and
\begin{equation}
\delta \theta (s)=\sum\limits_{n=0}^{\infty }\delta \theta _{n}\cos \left(
\frac{n\pi s}{L}\right) .  \label{deltatheta}
\end{equation}%
In terms of these modes, the bending energy of the chain is given by
\begin{equation}
\begin{array}{c}
E_{bend}=\frac{\kappa }{4L}\{-8\pi ^{2}\delta \theta
_{0}^{2}+\sum\limits_{n=1}^{\infty }(n^{2}-4)\pi ^{2}{\delta \theta _{n}}%
^{2}+\sum\limits_{n,odd}n^{2}\pi ^{2}\left( \delta \phi _{n}-\frac{8}{%
n^{2}\pi }\right) ^{2}+\sum\limits_{n,even}n^{2}\pi ^{2}\delta \phi
_{n}^{2}\}.%
\end{array}
\label{BendingEnergy}
\end{equation}%
In the above, the bending energy is independent of the modes $\delta \theta
_{2}$ and $\delta \phi _{0}$. These give two of the three rotational degrees
of freedom of the ring polymer in space. $\delta \phi _{0}$ corresponds to
the rotation about the $Z$ axis, while $\delta \theta _{2}$ corresponds to
rotation of the ring about X axis. A fluctuation of the form $\delta \theta
_{2s}\sin (2\pi s/L)$ leads to rotation about $Y$ axis. The value of $\delta
\theta _{2s}$ (the amount of of rotation contained) in an arbitrary $\delta
\theta (s)$ may be found from
\begin{equation}
\delta \theta _{2s}=\frac{2}{L}\int\limits_{0}^{L}\delta \theta (s)\sin
\left( \frac{2\pi s}{L}\right) .  \label{dtheta2s}
\end{equation}%
Using Eq. (\ref{deltatheta}) in Eq. (\ref{dtheta2s}), one gets%
\begin{equation}
\delta \theta _{2s}=2\sum\limits_{n}a_{n}\delta \theta _{n},
\end{equation}%
where
\begin{equation}
\begin{array}{c}
a_{n}=\frac{-4}{(n^{2}-4)\pi },\ \mathrm{\ if\ n\ is\ odd} \\
=0,\ \mathrm{\ if\ n\ is\ even}.%
\end{array}
\label{an}
\end{equation}%
While evaluating the partition function one must avoid integrating over the
rotational modes, since within our approximation scheme, these modes would
cause the partition function to diverge. One can remove these rotational
degrees of freedom by inserting the product of delta functions $\delta
(\delta \phi _{0})\delta (\delta \theta _{2})\delta (\delta \theta _{2s})$
to the functional integral and then taking the contribution of the
rotational modes to the partition function into account by explicitly
putting in the factor $8\pi ^{2}$. Then the probability for the loop
formation is given by
\begin{equation}
\begin{array}{c}
G(\mathbf{0,}L\mathbf{|0})=\frac{8\pi ^{2}}{Z_{R}}\int D\delta \theta
(s)\int D\delta \phi (s)\exp \left( -\frac{E_{bend}[\delta \theta (s),\delta
\phi (s)]}{k_{B}T}\right) \delta (\delta \phi _{0})\delta (\delta \theta
_{2})\delta (\delta \theta _{2s})\delta (\mathbf{R}).%
\end{array}
\label{pfthetaphi1}
\end{equation}%
$\mathbf{R}$ is the end to end vector for the polymer chain and $Z_{R}$ is
the partition function for the polymer, approximated by that appropriate for
a semi-flexible rod of length $L$ (see Eq. (\ref{Z_R})).

\subsection{The asymmetry in the three directions of motion at the ring
geometry}

\label{sectwo2}
In Ref. \cite{klssantoPRE2006}, we derived the expression
for the end-to-end vector $\mathbf{R}$, by expanding the components of $%
\mathbf{u}(s)$ as a Taylor series up to first order, which is
\begin{equation}
\mathbf{R}=L\left( \mathbf{i}\sum\limits_{n\ odd}^{\infty }a_{n}\delta \phi
_{n}-\mathbf{j}\frac{\delta \phi _{2}}{2}+\mathbf{k}\delta \theta
_{0}\right) .  \label{endtoendvector}
\end{equation}%
The components of ${\mathbf{R}}$ are given by, $R_{x}=L\sum_{n,odd}^{\infty
}a_{n}\delta \phi _{n}$, $R_{y}=-\frac{L}{2}\delta \phi _{2}$ and $%
R_{z}=L\delta \theta _{0}$. Thus $R_{x}$ can be changed by varying the value
of $\delta \phi _{n}$s for odd $n$. It can be easily seen from Eq. (\ref%
{BendingEnergy}), that increasing $\delta \phi _{n}$s with $n$ odd decreases
the bending energy of the chain towards a minimum at $\delta \phi
_{n}=8/n^{2}\pi $ ($n$ odd). Using this value for $\delta \phi _{n}$ ($n$
odd ) one gets
\begin{equation}
R_{x}=L\sum\limits_{n,odd}^{\infty }a_{n}\frac{8}{n^{2}\pi }=L,
\label{Rxrod}
\end{equation}%
since $\sum_{n,odd}^{\infty }a_{n}/n^{2}=\pi /8$. Hence, this value of $%
\delta \phi _{n}$ corresponds to a rod lying along the $X$ axis. Therefore,
the ring is unstable along $R_{x}$ and the bending energy along this
direction has the minimum at $R_{x}=L$. Also, increasing $\delta \theta _{0}$
decreases the bending energy and therefore, $R_{z}$ is also unstable. This
is because when $R_{z}$ is increased the ring changes into a helix, which
has less curvature and therefore less bending energy. (Note that we do not
take torsional energies into account in this analysis). But unlike $R_{x}=0$%
, $R_{z}=0$ is a maximum. It should be noted that $R_{z}=L$ corresponds to
the rod and should be a minimum, but our analysis does not reproduce this.
So, the instability along $R_{z}$ is only near the ring, where our analysis
is valid. Unlike $R_{x}$ and $R_{z}$, $R_{y}$ is stable, since the bending
energy of the ring increases when $\delta \phi _{2}$ is increased. Thus, the
motion in $R_{x}$, $R_{y}$ and $R_{z}$ directions are energetically
different.

\subsection{Bending energy of the rod}

\label{Bendingenergyrod} For a semiflexible chain, the minimum energy
configuration is the rod. On the unit sphere representing the tangents this
means that the random walker stays at the starting point. We take this point
to be $(\theta (s),\phi (s))=(\pi /2,0)$, which corresponds to the rod lying
along the X axis. Unlike the great circle, the straight rod is an exact
minimum energy configuration. In this case, the fluctuations can be
incorporated by letting
\begin{equation}
\lbrack \theta (s),\phi (s)]=\left[ \frac{\pi }{2}+\delta \theta (s),\delta
\phi (s)\right] .  \label{urod}
\end{equation}%
The bending energy of the rod correct up to the second order in fluctuations
is then given by
\begin{equation}
E_{rod}=\frac{\kappa }{2}\int\limits_{0}^{L}\left\{ \left( \frac{d\delta
\theta (s)}{ds}\right) ^{2}+\left( \frac{\delta \phi (s)}{ds}\right)
^{2}\right\} ds.  \label{bendrod}
\end{equation}%
Using the expansions Eq. (\ref{deltaphi}) and Eq. (\ref{deltatheta}), one
gets
\begin{equation}
E_{rod}=\frac{\kappa }{4L}\sum\limits_{n=0}^{2\mathcal{N}}n^{2}\pi
^{2}(\delta \theta _{n}^{2}+\delta \phi _{n}^{2}).  \label{Erod}
\end{equation}%
Unlike the ring, the rod has only two rotational degrees of freedom. From
Eq. (\ref{Erod}), it follows that these are the modes $\delta \phi _{0}$ and
$\delta \theta _{0}$. For the convenience of bookkeeping, we assume that the
number of $\delta \phi _{n}$ and $\delta \theta _{n}$ modes and both equal
to $2\mathcal{N}$, with $n=0,1,2...(2\mathcal{N}-1)$. \ Thus there are $4%
\mathcal{N}$ modes in total, with $\mathcal{N}\rightarrow \infty $.

\section{The normal coordinates and the Green's function}

\label{NormalCoordinates} In this section, we use the approximation scheme
for the bending energy described in Sec. \ref{BendingenergyLoop}, to analyze
the dynamics of loop formation. The approximation of Eq. (\ref{BendingEnergy}%
) is valid only near the most important path corresponding to the loop,
since the fluctuations about this path are assumed to be small. The time
evolution of the chain may be described by the multidimensional Green's
function $G_{0}(\mathbf{\Psi },t|\mathbf{\Psi }_{0})$, where $\mathbf{\Psi }%
^{\dag }=(\mathbf{\Phi }^{\dag },\mathbf{\Theta }^{\dag })$ with $\mathbf{%
\Phi }^{\dag }=(\delta \phi _{0},\delta \phi _{2},..\delta \phi _{2\mathcal{N%
}-2},\delta \phi _{1},\delta \phi _{3},..\delta \phi _{2\mathcal{N}-1})$
(note that we have separated out the even and odd modes) and $\mathbf{\Theta
}^{\dag }=(\delta \theta _{0},\delta \theta _{2},..\delta \theta _{2\mathcal{%
N}-2},\delta \theta _{1},\delta \theta _{3},..\delta \theta _{2\mathcal{N}%
-1})$. The superscript $\dagger $ stands for transpose. \ The bending energy
of the polymer near the ring configuration is given by Eq. (\ref%
{BendingEnergy}) and therefore, $G_{0}$ for configurations close to the ring
may be obtained by solving the corresponding equations of motion. Because of
our approximation for the energy, the Green's function so obtained is not
valid for large $\mathbf{R}$. Yet, the sink-sink correlation function, $%
\mathcal{D}(t)$ of Eq. (\ref{sinkcor}), may still be evaluated, since the
sink function $\mathcal{S}(\mathbf{R})$ is nonzero only for very small
values of $\mathbf{R}$. This of course, is approximate. The function $G_{0}$
may be found by solving the equations of motion of the polymer near the loop
configuration. The angle coordinates, $\mathbf{\Psi }$ are not normal
coordinates, since the kinetic energy of the chain has terms that couple
these (see Appendix \ref{AppendixKEMatrices}). As a result, the equations of
motion of the chain in terms of them are coupled. This coupling may be
avoided by working with the normal modes, which may be found by solving the
corresponding eigenvalue problem (Eq. (\ref{DiagonalizeVeff})). Then the
dynamics of the chain can be reduced to the dynamics of a particle in a
multidimensional harmonic potential. The Green's function $G_{0}$ is
obtained as a product of the one-dimensional Green's functions corresponding
to each of the normal modes.

\subsection{The Hamiltonian and the normal modes}

The kinetic energy of the polymer in the center of mass frame is
\begin{equation}
T=\frac{\rho }{2}\int\limits_{0}^{L}\left[ \frac{\partial \mathbf{r}(s)}{%
\partial t}\right] ^{2}dt,  \label{kepoly}
\end{equation}%
where $\mathbf{r}(s)$ is given by Eq. (\ref{positionvector}). Near the rod
ring configuration, the kinetic energy of the polymer in terms of the
Fourier modes $\delta \theta _{n}$ and $\delta \phi _{n}$ may be written as
\begin{equation}
\begin{array}{c}
T_{R}=\frac{\rho L^{3}}{2}\overset{.}{\mathbf{\Psi }}_{R}^{\dag }\mathcal{T}%
_{R}\overset{.}{\mathbf{\Psi }_{R}}.%
\end{array}
\label{keRod}
\end{equation}%
The dot in $\overset{.}{\mathbf{\Psi }_{R}}$ represents differentiation with
respect to time. The subscripts $R(L)$ in $\mathbf{\Psi }_{R}$($\mathbf{\Psi
}_{L}$) are used to indicate that these are deviations measured from values
appropriate for the rod (loop) geometry. $\ \mathcal{T}_{R}\mathbf{\ }$is
the kinetic energy matrix appropriate near the rod configuration. \ It has a
block diagonal structure, having no matrices connecting the $\theta $ and $%
\phi $ modes. \ Even within the $\theta $ and $\phi $ modes, odd and even
modes are decoupled. \ Hence $\mathcal{T}_{R}$ may be written as
\begin{equation*}
\mathcal{T}_{R}=\left[
\begin{array}{cccc}
\mathbf{T}_{R}^{\phi e} & \mathbf{0} & \mathbf{0} & \mathbf{0} \\
\mathbf{0} & \mathbf{T}_{R}^{\phi o} & \mathbf{0} & \mathbf{0} \\
\mathbf{0} & \mathbf{0} & \mathbf{T}_{R}^{\theta e} & \mathbf{0} \\
\mathbf{0} & \mathbf{0} & \mathbf{0} & \mathbf{T}_{R}^{\theta e}%
\end{array}%
\right] .
\end{equation*}%
Detailed structures of the $\mathbf{T}$ matrices is given in Appendix \ref%
{AppendixKEMatrices}. In a similar fashion, near the loop configuration the
kinetic energy is given by
\begin{equation}
\begin{array}{c}
T_{L}=\frac{\rho L^{3}}{2}\overset{.}{\mathbf{\Psi }}_{L}^{\dag }\mathcal{T}%
_{L}\overset{.}{\mathbf{\Psi }_{L}}.%
\end{array}
\label{keLoop}
\end{equation}%
The angles $\overset{.}{\mathbf{\Psi }}_{L}^{\dag }=$ $(\mathbf{\Phi }%
_{L}^{\dag },\mathbf{\Theta }^{\dag })$ with $\mathbf{\Phi }_{L}^{\dag
}=(\delta \phi _{0},\delta \phi _{2},..\delta \phi _{2\mathcal{N}-2},\delta
\phi _{1}^{^{\prime }},\delta \phi _{3}^{^{\prime }},..\delta \phi _{2%
\mathcal{N}-1}^{\prime })$, where $\delta \phi _{n}^{^{\prime }}=\delta \phi
_{n}-8/(n^{2}\pi ).$ \ Like $\mathcal{T}_{R},$ $\mathcal{T}_{L}$ too has a
block diagonal structure, with the blocks given by the matrices $\mathbf{T}%
_{L}^{\phi e},\mathbf{T}_{L}^{\phi o},\mathbf{T}_{L}^{\theta e}$ and $%
\mathbf{T}_{L}^{\theta o}$. \ The forms of these too are given in Appendix %
\ref{AppendixKEMatrices}. Note that these matrices have no length ($L)$
dependence. It is found that (see Appendix \ref{AppendixKEMatrices}) the
modes of odd and even $n$ decouple. One may rewrite Eq. (\ref{BendingEnergy}%
) as
\begin{equation}
\begin{array}{c}
E_{bend}=\frac{\kappa }{4L}\left[ -8\pi ^{2}\delta \theta
_{0}^{2}+\sum\limits_{n\ even}(n^{2}-4)\pi ^{2}{\delta \theta _{n}}^{2}%
\right] + \\
\frac{\kappa }{4L}\left[ \sum\limits_{n,odd}n^{2}\pi ^{2}\left( \delta \phi
_{n}^{\prime }\right) ^{2}+\sum\limits_{n,even}n^{2}\pi ^{2}\delta \phi
_{n}^{2}\right] .%
\end{array}
\label{Ebenda}
\end{equation}%
or as%
\begin{equation}
E_{bend}=\frac{\kappa }{2L}\mathbf{\Psi }_{L}^{\dag }\mathcal{V}_{L}\mathbf{%
\Psi }_{L}.  \label{EbendMatrixForm}
\end{equation}%
Thus the total energy of the polymer molecule near the loop configuration is
\begin{equation}
E=\frac{\rho L^{3}}{2}\overset{.}{\mathbf{\Psi }}_{L}^{\dag }\mathcal{T}_{L}%
\overset{.}{\mathbf{\Psi }_{L}}+\frac{\kappa }{2L}\mathbf{\Psi }_{L}^{\dag }%
\mathcal{V}_{L}\mathbf{\Psi }_{L}.  \label{LoopTotalEnergy}
\end{equation}%
Like $\mathcal{T}_{L}$, $\mathcal{V}_{L}$ too are block diagonal. \ The
matrices of which $\mathcal{V}_{L}$ is composed of are $\mathbf{V}_{L}^{\phi
e},\mathbf{V}_{L}^{\phi o},\mathbf{V}_{L}^{\theta e}$ and $\mathbf{V}%
_{L}^{\theta o}.$ \ Each one of them is diagonal and have matrix elements
given by $(\mathbf{V}_{L}^{\phi o})_{nm}=\delta _{mn}n^{2}\pi ^{2}/2=(%
\mathbf{V}_{L}^{\phi e})_{nm}$; $(\mathbf{V}_{1}^{\theta })_{00}=-4\pi ^{2}$
and all other matrix elements being given by $(\mathbf{V}_{L}^{\theta
})_{nm}=\delta _{mn}(n^{2}-4)\pi ^{2}/2$.

We will choose the sink function $\mathcal{S}(\mathbf{r})$ as a function
only of the end-to-end vector $\mathbf{R}$ (see next section). The dynamics
of the closing process must be unaffected by the spatial rotations of the
polymer. Hence, the sink-sink correlation function, $\mathcal{D}(t)$ of Eq. (%
\ref{sinkcor}) is independent of the rotational modes $\delta \phi
_{0},\delta \theta _{2}$ and $\delta \theta _{2s}$. From Eq. (\ref%
{endtoendvector}) it follows that $\delta \theta _{n}$ modes with odd $n$ do
not contribute to the end-to-end separation $\mathbf{R}$ of the polymer. As $%
\mathbf{R}$ has no dependence on the odd $\delta \theta _{n}$ modes, they
are irrelevant for the dynamics of closing process.

We define $\mathbf{Y}$ by
\begin{equation}
\mathcal{T}^{1/2}\mathbf{\Psi }_{L}=\mathcal{U}\mathbf{Y,}  \label{DefiningY}
\end{equation}%
where $\mathcal{U}$ is to be defined below. Then the energy becomes
\begin{equation}
E=\frac{\rho L^{3}}{2}\overset{.}{\mathbf{Y}}^{\dag }\mathcal{U^{\dag }U}%
\overset{.}{\mathbf{Y}}+\frac{\kappa }{2L}\mathbf{Y}^{\dag }\mathcal{U^{\dag
}\mathcal{T}}^{-1/2}\mathcal{\mathcal{V}}_{L}\mathcal{\mathcal{\mathcal{T}}}%
^{-1/2}\mathcal{U}\mathbf{Y.}  \label{cond1}
\end{equation}%
Taking $\mathcal{U}$ to be a unitary matrix, which diagonalizes $\mathcal{%
\mathcal{T}}^{-1/2}\mathcal{\mathcal{V}}_{L}\mathcal{\mathcal{\mathcal{T}}}$
$^{-1/2}$ to give the diagonal matrix $\mathcal{K}$, as
\begin{equation}
\mathcal{U^{\dag }\mathcal{T}}^{-1/2}\mathcal{\mathcal{V}}_{L}\mathcal{%
\mathcal{\mathcal{T}}}^{-1/2}\mathcal{U}=\mathcal{K}  \label{DiagonalizeVeff}
\end{equation}%
we get
\begin{equation}
E=\frac{\rho L^{3}}{2}\overset{.}{\mathbf{Y}}^{\dag }\overset{.}{\mathbf{Y}}+%
\frac{\kappa }{2L}\mathbf{Y}^{\dag }\mathcal{K}\mathbf{Y}
\label{LoopTotalEnergyNormalModes}
\end{equation}%
with
\begin{equation}
\mathcal{K}=\left[
\begin{array}{cccc}
\mathbf{K}^{\phi e} & \mathbf{0} & \mathbf{0} & \mathbf{0} \\
\mathbf{0} & \mathbf{K}^{\phi o} & \mathbf{0} & \mathbf{0} \\
\mathbf{0} & \mathbf{0} & \mathbf{K}^{\theta e} & \mathbf{0} \\
\mathbf{0} & \mathbf{0} & \mathbf{0} & \mathbf{K}^{\theta o}%
\end{array}%
\right] ,  \label{DiagonalK}
\end{equation}%
The block diagonal structures of $\mathcal{T}$ and $\mathcal{V}$ imply that $%
\mathcal{U}$ also has a block diagonal structure, with matrices $\mathbf{U}%
^{\phi e},\mathbf{U}^{\phi o},\mathbf{U}^{\theta e}$ and $\mathbf{U}^{\theta
o}$ occuring along the diagonal. \ The energy may be written in terms of the
components of $\mathbf{Y}$ and $\mathcal{K}$ as
\begin{equation}
E_{close}=\frac{\rho L^{3}}{2}\sum\limits_{n=1}^{4\mathcal{N}}\dot{Y}%
_{n}^{2}+\frac{\kappa }{2L}\sum\limits_{n=1}^{4\mathcal{N}}k_{n}Y_{n}^{2}.
\label{EnergyNormalModes}
\end{equation}%
Note that $k_{n}$ are not dependent on the length $L$ of the chain. Of the
modes $\mathbf{Y,}$ $Y_{n},$ with $n=3\mathcal{N}+1$ to $4\mathcal{N}$ arise
from odd $\delta \theta _{n}.$ \ The end to end vector $\mathbf{R}$ has no
dependence on them. \ Hence these $Y_{n}$ play no role in the dynamics of
loop formation, occuring near the loop geometry. \ Therefore, we focus on
the remaining modes. We write the remaining normal coordinates $\mathbf{Y}%
_{I}\mathbf{=}(Y_{1},Y_{2},.....Y_{3\mathcal{N}})$ as
\begin{equation}
\mathbf{Y}_{I}=(\mathbf{x},\mathbf{y},\mathbf{z}).  \label{Definexyzvectors}
\end{equation}%
i.e, $\mathbf{Y}_{I}=(x_{1},x_{2},..x_{\mathcal{N}},y_{1},y_{2},....y_{%
\mathcal{N}},z_{1},z_{2},....z_{\mathcal{N}})$. $\ \ $Within $\mathbf{x}$ ($%
\mathbf{y}$ or $\mathbf{z),}$ we take the modes to be arranged in the order
of increasing eigenvalues, and we label them as $k_{xn}$ ($k_{yn}$ or $%
k_{zn} $), with $n$ varying from $1$ to $\mathcal{N}$. $x_{n}$ are the
normal modes corresponding to the modes the $\delta \phi _{n}^{\prime }$ and
these are all stable modes, as may be inferred by looking at the expression
for energy of Eq. (\ref{BendingEnergy}). \ $y_{n}$ corresponds to the even $%
\delta \phi _{n}$s. \ $\delta \phi _{0}$ is a rotational mode and
correspondingly, $k_{y1}=0$. \ Of the even $\delta \theta _{n}$ modes, one
is unstable, viz., the one that corresponds to $\delta \theta _{0}$. It
leads to separation between the two ends in the Z-direction and is unstable,
as we already discussed. $\delta \theta _{2}$ is a rotational mode and would
give us a zero eigenvalue. \ Thus we have $k_{z1}$ negative and $k_{z2}$
equal to zero. \ Note that the eigenvalues $k_{n}$ have no dependence on $%
\kappa $ or $L$ and hence are universal numbers. \ Their values up to $n=10$
are given in Table \ref{table1}.

The end-to-end distance $\mathbf{R}$ may be expressed in terms of the normal
coordinates $\mathbf{Y}_{I}$. The $x$-component
\begin{equation}
R_{x}=L\sum\limits_{n}a_{n}\delta \phi _{n}=L\sum\limits_{n}a_{n}\delta \phi
_{n}^{\prime }+\frac{8L}{\pi }\sum\limits_{n}\frac{a_{n}}{n^{2}},
\label{Rxintermsofphiodd}
\end{equation}%
which on using $a_{n}=-4/(n^{2}-4)\pi $ ($n$ odd) becomes
\begin{equation}
R_{x}=L\sum\limits_{n\ odd}a_{n}\delta \phi _{n}^{\prime }+L.
\label{Rxintermsofphiprime}
\end{equation}%
Now, in terms of the normal coordinates (i.e., inverting Eq. ((\ref%
{DefiningY})), $\delta \phi _{n}^{\prime }=\sum\limits_{m=1}^{\mathcal{N}}%
\left[ \left( \mathbf{T}^{\phi o}\right) ^{-1/2}\mathbf{U}^{\phi o}\right]
_{nm}x_{m}$. Hence,
\begin{equation}
R_{x}-L=L\sum\limits_{n=1}^{\mathcal{N}}f_{n}x_{n},  \label{Rxintermsofx}
\end{equation}%
with%
\begin{equation}
f_{n}=\sum\limits_{k=1}^{\mathcal{N}}a_{2k-1}\left[ \left( \mathbf{T}^{\phi
o}\right) ^{-1/2}\mathbf{U}^{\phi o}\right] _{kn}.  \label{Definitionoffn}
\end{equation}%
Similarly,
\begin{equation}
R_{y}=L\sum\limits_{n=1}^{\mathcal{N}}g_{n}y_{n},  \label{Ryintermsofy}
\end{equation}%
where
\begin{equation}
g_{n}=\left[ \left( \mathbf{T}^{\phi e}\right) ^{-1/2}\mathbf{U}^{\phi e}%
\right] _{2n}/2.  \label{Definitionofgn}
\end{equation}%
Note that $y_{1}$ is a rotational mode and the corresponding $g_{1}$ would
be zero. Therefore, the above sum may be modified to
\begin{equation}
R_{y}=L\sum\limits_{n\neq 2}^{\mathcal{N}}g_{n}y_{n}.
\label{Ryintermsofyfinal}
\end{equation}%
The $z$ component
\begin{equation}
R_{z}=L\sum\limits_{n=1}^{\mathcal{N}}h_{n}z_{n},  \label{Rzn}
\end{equation}%
with
\begin{equation}
h_{n}=\left[ \left( \mathbf{T}^{\theta e}\right) ^{-1/2}\mathbf{U}^{\theta e}%
\right] _{1n}.  \label{Definitionofhn}
\end{equation}%
Again, $z_{2}$ being the rotational mode, this may be written as
\begin{equation}
R_{z}=L\sum\limits_{n\neq 2}^{\mathcal{N}}h_{n}z_{n}.  \label{Rzintermsofz}
\end{equation}

\begin{table}[h]
\centering%
\begin{tabular}{|c|c|c|c|c|c|c|}
\hline
$n$ & $k_{xn} $ & $f_n^2/k_{xn} $ & $k_{yn} $ & $g_n^2/k_{yn} $ & $k_{zn} $
& $h_n^2/k_{zn} $ \\ \hline
1 & $298.54 $ & $3.7839E-02 $ & $0.0 $ & $0.0 $ & $-1241.1 $ & $-2.5996E-02 $
\\
2 & $7.1266E+03 $ & $1.5524E-04 $ & $1.4131E+03 $ & $1.1665E-02 $ & $0.0 $ &
$0.0 $ \\
3 & $6.3197E+04 $ & $1.4907E-06 $ & $2.4549E+04 $ & $7.9245E-04 $ & $%
2.8705E+04 $ & $5.0074E-04 $ \\
4 & $2.5382E+05 $ & $6.3190E-08 $ & $1.3492E+05 $ & $1.3331E-04 $ & $%
1.5259E+05 $ & $1.0042E-04 $ \\
5 & $7.0125E+05 $ & $5.8587E-09 $ & $4.3631E+05 $ & $3.9529E-05 $ & $%
4.7411E+05 $ & $3.2983E-05 $ \\
6 & $1.5655E+06 $ & $8.6906E-10 $ & $1.0696E+06 $ & $1.5805E-05 $ & $%
1.1339E+06 $ & $1.3963E-05 $ \\
7 & $3.0455E+06 $ & $1.7658E-10 $ & $2.2137E+06 $ & $7.5909E-06 $ & $%
2.3108E+06 $ & $6.9411E-06 $ \\
8 & $5.3826E+06 $ & $4.5142E-11 $ & $4.0872E+06 $ & $4.1420E-06 $ & $%
4.2234E+06 $ & $3.8728E-06 $ \\
9 & $8.8695E+06 $ & $1.3891E-11 $ & $6.9516E+06 $ & $2.4991E-06 $ & $%
7.1330E+06 $ & $2.3728E-06 $ \\
10 & $1.3893E+07 $ & $5.1570E-12 $ & $1.1123E+07 $ & $1.6576E-06 $ & $%
1.1355E+07 $ & $1.5914E-06 $ \\ \hline
\end{tabular}%
\caption{The dimensionless eigenvalues $k_{n}$ and the corresponding values
of $f_n^2/k_{xn} $ , $g_n^2/k_{yn} $ and $h_n^2/k_{zn}$. }
\label{table1}
\end{table}

\subsection{The equations of motion and the Green's function}

\subsubsection{The equations of motion}

The equation of motion of the polymer in a dissipative environment is given
by
\begin{equation}
\rho \ddot{\mathbf{r}}+\rho \gamma \dot{\mathbf{r}}+\frac{\delta E[\mathbf{r}%
(s)]}{\delta {\mathbf{r}}(s)}=\zeta (s,t),  \label{lgvn1}
\end{equation}%
where $E[\mathbf{r}(s)]$ is the energy functional of the chain and $\zeta
(s,t)$ is the stochastic force acting on the $s^{th}$ segment of the chain. $%
\zeta (s,t)$ is assumed to obey $<\zeta (s,t)>=0$ and $<\zeta (s,t)\zeta
(s^{\prime },t^{\prime })>=2k_{B}T\rho \gamma \delta (t-t^{\prime })\delta
(s-s^{\prime })$. In the over-damped limit, one may write Eq. (\ref{lgvn1})
as
\begin{equation}
\rho \gamma \dot{\mathbf{r}}+\frac{\delta E}{\delta {\mathbf{r}}(s)}=\zeta
(s,t).  \label{eqmoverdamped}
\end{equation}%
Through the use of a system-plus-reservoir model, this equation can be
equivalently expressed in the angle coordinates as (see Ref. \cite%
{klssantoPRE2006})
\begin{equation}
\rho L^{3}\gamma \sum\limits_{n}T_{mn}^{\phi }\delta \dot{\phi}_{n}+\frac{%
\kappa }{L}\sum_{n}V_{mn}^{\phi }\delta \phi _{n}=\zeta _{m}^{\phi }(t)
\label{eqmphi1}
\end{equation}%
and
\begin{equation}
\rho L^{3}\gamma \sum\limits_{n}T_{mn}^{\theta }\delta \dot{\theta}_{n}+%
\frac{\kappa }{L}\sum_{n}V_{mn}^{\phi }\delta \theta _{n}=\zeta _{m}^{\theta
}(t).  \label{eqmtheta2}
\end{equation}%
Equations (\ref{eqmphi1}) and (\ref{eqmtheta2}) represent sets of coupled
first order differential equations. For the ring, we can express them in
terms of the normal modes. In terms of the normal modes $Y_{n}$, equations (%
\ref{eqmphi1}) and (\ref{eqmtheta2}) represent a set of independent one
dimensional Langevin equations,
\begin{equation}
\dot{Y}_{n}+\frac{\kappa k_{n}}{\rho L^{4}\gamma }Y_{n}=\zeta _{n}(t),
\label{lneq}
\end{equation}%
where $\zeta _{n}(t)$, is a white Gaussian noise with $<\zeta _{n}(t)>=0$
and $<\zeta _{n}(t)\zeta _{m}(t^{\prime })>=2k_{B}T\gamma/(\rho
L^{3})\delta (t-t^{\prime })\delta _{mn}$. From Eq. (\ref{lneq}), it follows
that the relaxation time of each mode $\sim L^{4}$.

\subsubsection{The Green's function}

Eq. (\ref{lneq}) describes a particle of mass $\rho L^{3}$, subject to
friction $\gamma $ in a one-dimensional harmonic potential, $\kappa
k_{n}Y_{n}^{2}/2L$. The Green's function for it is given by \cite%
{gardiner1983,RiskenFP1989}
\begin{equation}
\begin{array}{c}
G_{n}(Y_{n},Y_{n}^{\prime };t)=\left\{ \frac{2\pi Lk_{B}T}{\kappa k_{n}}%
[1-\exp (-2t/\tau _{n})]\right\} ^{-1/2} \\
\exp \left\{ -\frac{\kappa k_{n}(Y_{n}-\exp (-t/\tau _{n})Y_{n}^{\prime
})^{2}}{2Lk_{B}T[1-\exp (-2t/\tau _{n})]}\right\}%
\end{array}
\label{Green'sfunctionforYn}
\end{equation}%
with $\tau _{n}=\tau _{0}/k_{n}$ with $\tau _{0}=\rho L^{4}\gamma /\kappa $.
$G_{n}$ is the conditional probability to find the particle at $Y_{n}$ at
time $t$, given that it was at $Y_{n}^{\prime }$ at $t=0$. Eq. (\ref%
{Green'sfunctionforYn}) is valid for both positive and negative $k_{n}$ \cite%
{RiskenFP1989}. \ \bigskip

\section{The equilibrium distribution}

\label{EquilibriumDistribution} In this section, we derive an approximate
equilibrium distribution function for the semiflexible polymer near the loop
configuration in terms of the angle coordinates. The partition function of
the polymer is
\begin{equation}
Z=\int d\mathbf{r}\int d\mathbf{p}\exp (-\beta H[\mathbf{r},\mathbf{p}]).
\label{Z}
\end{equation}%
In terms of the angle coordinates, the partition function is given by
\begin{equation}
Z=\int d\mathbf{\Psi }\int d\mathbf{p_{\mathbf{\Psi }}}\exp (-\beta H[%
\mathbf{\Psi },\mathbf{p_{\mathbf{\Psi }}}]),  \label{Zintermsofangle}
\end{equation}%
where $\mathbf{p}_{\mathbf{\Psi }}$ are the momenta conjugate to the angle
coordinates $\mathbf{\Psi }$. In the integral, the configurations that
contribute the most are the ones near the rod configuration. For such
configurations, the energy is given by (see Eq. (\ref{Erod}))
\begin{equation}
E=\frac{\rho L^{3}}{2}\overset{.}{\mathbf{\Psi }}_{R}^{\dag }\mathcal{T}_{R}%
\overset{.}{\mathbf{\Psi }_{R}}+\frac{\kappa }{2L}\mathbf{\Psi }_{R}^{\dag }%
\mathcal{V}_{R}\mathbf{\Psi }_{R},  \label{EnergyRod}
\end{equation}%
where $\mathcal{V}_{R}$ too is block diagonal, consisting of
\begin{equation}
\mathcal{V}_{R}=\left[
\begin{array}{cccc}
\mathbf{V}_{R}^{\phi e} & \mathbf{0} & \mathbf{0} & \mathbf{0} \\
\mathbf{0} & \mathbf{V}_{R}^{\phi o} & \mathbf{0} & \mathbf{0} \\
\mathbf{0} & \mathbf{0} & \mathbf{V}_{R}^{\theta e} & \mathbf{0} \\
\mathbf{0} & \mathbf{0} & \mathbf{0} & \mathbf{V}_{R}^{\theta e}%
\end{array}%
\right] .  \label{VRod}
\end{equation}%
The matrix elements of each of the matrices on the right hand are given by $%
\left( \mathbf{V}_{R}^{\phi e}\right) _{mn}=\frac{n^{2}\pi ^{2}}{2}\delta
_{mn}=\left( \mathbf{V}_{R}^{\theta e}\right) _{mn}$ and $\left( \mathbf{V}%
_{R}^{\phi o}\right) _{mn}=\frac{n^{2}\pi ^{2}}{2}\delta _{mn}=\left(
\mathbf{V}_{R}^{\theta o}\right) _{mn}$. The momenta conjugate to $\mathbf{%
\Psi }_{R}$ is $\mathbf{p}_{\mathbf{\Psi }_{R}}=\rho L^{3}\mathcal{T}%
_{R}^{-1}\overset{.}{\overset{.}{\mathbf{\Psi }}_{R}}$ and hence the
Hamiltonian is given by
\begin{equation}
H_{R}=\frac{1}{2\rho L^{3}}\mathbf{p}_{\mathbf{\Psi }}^{\dagger }\mathcal{T}%
_{R}^{-1}\mathbf{p}_{\mathbf{\Psi }}+\frac{\kappa }{2L}\mathbf{\Psi }%
_{R}^{\dagger }\mathcal{V}_{R}\mathbf{\Psi }_{R}.  \label{HamiltonianRod}
\end{equation}%
The partition function of the rod can be evaluated now, using equatioins (%
\ref{EnergyRod}) and (\ref{HamiltonianRod}). \ polymer given in Eq. (\ref%
{Zintermsofangle}) near the rod can be evaluated now. \ The rod has two
rotational modes, which are $\delta \phi _{0}$ and $\delta \theta _{0}$. \
Integrating over them would give a factor of $4\pi $. \ Performing the
integration over the remaining $\delta \phi _{n}$ and $\delta \theta _{n}$
and integrating over all the momenta, give
\begin{eqnarray}
Z_{R} &=&4\pi (2\pi \rho L^{3}k_{B}T)^{2\mathcal{N}}(\det \mathcal{T}%
_{R})^{1/2}(2\pi k_{B}TL/\kappa )^{2\mathcal{N}-1}({\det }^{\prime }\mathcal{%
V}_{R})^{-1/2}.  \label{Z_R} \\
&=&\frac{1}{((2\mathcal{N}-1)!)^{2}}2^{6\mathcal{N}}L^{8\mathcal{N}-1}\pi
^{2}\beta ^{1-4\mathcal{N}}\kappa ^{1-2\mathcal{N}}\rho \ ^{2\mathcal{N}}%
\sqrt{\left\vert \mathbf{T}_{R}^{\theta }\right\vert }\sqrt{\left\vert
\mathbf{T}_{R}^{\phi }\right\vert }
\end{eqnarray}%
The prime on the determinant in ${\det }^{\prime }\mathcal{V}_{R}$ indicates
that the zero eigenvalues (rotational modes) are excluded. Note that we use $%
|\mathbf{T}_{R}^{\phi }|$ to denote $|\mathbf{T}_{R}^{\phi e}||\mathbf{T}%
_{R}^{\phi o}|$ and $|\mathbf{T}_{R}^{\theta }|$ to denote $|\mathbf{T}%
_{R}^{\theta e}||\mathbf{T}_{R}^{\theta o}|$. For configurations close to
the loop, the Hamiltonian can be approximated by
\begin{equation}
H_{L}=\frac{1}{2\rho L^{3}}\mathbf{p}_{\mathbf{\Psi }_{L}}^{\dagger }%
\mathcal{T}_{L}^{-1}\mathbf{p}_{\mathbf{\Psi }_{L}}+\frac{\kappa }{2L}%
\mathbf{\Psi }_{L}^{\dagger }\mathcal{V}_{L}\mathbf{\Psi }_{L}
\label{HamiltonianLoop}
\end{equation}%
This can be used to calculate the equilibrium distribution near the loop
conformation. \ In particular, we are interested in\ the probability of
contact between the two ends at equilibrium $G(\mathbf{0,}L\mathbf{)}$
defined by
\begin{equation}
G(\mathbf{0,}L\mathbf{)=}\frac{1}{Z}\int d\mathbf{p}_{\mathbf{\Psi }}\int d%
\mathbf{\Psi }\exp (-\beta H)\delta (\mathbf{R)}.  \label{G(0,L)}
\end{equation}%
$\delta (\mathbf{R)}$ in the above ensures that the two ends of the chain
are in contact. \ Our strategy in the calculation is as follows. \ The major
contribution to the partition function comes from rod like conformations. \
Hence we approximate $Z\cong $ $Z_{R}$. Near the loop geometry, we can use
the approximation $H\cong $ $H_{L}$ and perform the integrals over the
angles, with rotational degrees of freedom easily accounted for. Thus \
\begin{equation}
G(\mathbf{0},L)=\frac{1}{Z_{R}}\int d\mathbf{p}_{\mathbf{\Psi }_{L}}\int d%
\mathbf{\Psi }_{L}\exp (-\beta H_{L})\delta (\mathbf{R)}.
\label{G(0,L)approximate}
\end{equation}%
The integrals over momenta are easy to perform.\ There are three rotational
degrees of freedom, which can be removed by inserting $\delta (\delta \phi
_{0})\delta (\delta \theta _{2c})\delta (\delta \theta _{2s})$ into the
integrand, and their contribution accounted by introducing \ a
multiplicative factor of $8\pi ^{2}$. \ Then the integrations can be
performed, one by one, after using the integral representation for $\delta (%
\mathbf{R)}$ and using Eq. (\ref{endtoendvector}) for $\mathbf{R}$. \ Thus%
\begin{equation}
G(\mathbf{0},L)=\frac{8\pi ^{2}}{Z_{R}}\int d\mathbf{p}_{\mathbf{\Psi }%
_{L}}\int d\mathbf{\Psi }_{L}\exp (-\beta H_{L})\delta (\mathbf{R)}\delta
(\delta \phi _{0})\delta (\delta \theta _{2c})\delta (\delta \theta _{2s}).
\label{G(0,L,0)}
\end{equation}%
The details the calculation are given in the Appendix \ref{EvaluationofG}
and the result is
\begin{equation}
G(\mathbf{0,}L)=\frac{16\sqrt{2}e^{-\frac{4\pi ^{2}\beta \kappa }{3L}}\pi
^{3}\beta ^{2}\kappa ^{2}\sqrt{\left\vert \mathbf{T}_{L}^{\phi }\right\vert }%
}{3L^{5}\sqrt{\left\vert \mathbf{T}_{R}^{\phi }\right\vert }}
\label{G(0,L)Answer}
\end{equation}%
with $\left\vert \mathbf{T}_{L}^{\phi }\right\vert =\det (\mathbf{T}%
_{L}^{\phi }).$ Putting in numerical values (see Eq. (\ref{TRphiTLphiratio}%
)), we find
\begin{equation}
G(\mathbf{0,}L)=1522.06e^{-\frac{4\pi ^{2}\beta \kappa }{3L}}\frac{\beta
^{2}\kappa ^{2}}{L^{5}}  \label{Gour}
\end{equation}%
\begin{figure}[tbp]
\includegraphics[width=0.5\linewidth]{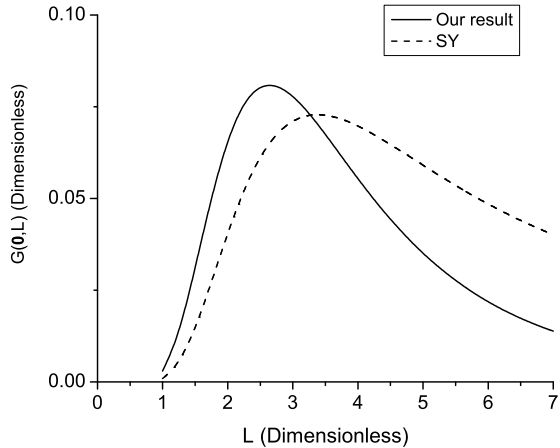}\newline
\caption{Comparison of our result for $G(\mathbf{0},L)$ (full line) with the
result of Shimada and Yamakawa\protect\cite{shimadaJCP1984} (dashed line).}
\label{compare}
\end{figure}
a form that is in agreement with the results of Shimada and Yamkawa \cite%
{shimadaJCP1984}. \ Note that the persistence length of the chain, $%
l_{p}=\beta \kappa $.\ We give a compartive plot of the our function and
their function $G_{SY}(\mathbf{0,}L)$ ($=896.32e^{-\frac{14.054l_{p}}{L}%
+0.246L/l_{p}}(l_{p}/L)^{5}$) in Fig. \ref{compare}. It is clear that there
is fair agreement between the two. The value of $L$ at which the maximum
occurs is $L=2.63l_{p}$ in our $G(\mathbf{0,}L)$, while it occurs at $%
L=3.37l_{p}$ for the results of Shimada and Yamakawa \cite{shimadaJCP1984}.

It is interesting to ask how the $L^{-5}$ term \ in Eq. (\ref{Gour}) comes
about. \ \ The Dirac delta function $\delta (\mathbf{R)}$ contributes
\thinspace $L^{-3}$. \ $G(\mathbf{0},L)$ would have the ratio of the
partition function for the loop conformation to that of the rod, and this
contributes a factor of $\left( \beta \kappa /L\right) ^{1/2}$. \ Further,
the fact that the potential energy depends on $\kappa /L$ term causes three
factors of $\left( \beta \kappa /L\right) ^{1/2}$ (from the three components
of $\mathbf{R}$ contained in the probability density at $R_{i}=0,$ with $%
i=x,y$ or $z$ and comes from the fact that the larger the value of $L$, the
broader the distribution of the $R_{i}$). \ These multiply together give the
factor $\beta ^{2}\kappa ^{2}/L^{5}$ in Eq. (\ref{Gour}).

\section{The time for loop formation}

\label{ClosingTime} We now evaluate the average loop formation time using
the approach outlined in Section \ref{Closure}. \ The quantity $v_{eq}$ is
\begin{equation}
v_{eq}=\left\langle \mathcal{S}(\mathbf{R})\right\rangle  \label{veq=S(R)}
\end{equation}%
with $\left\langle S(\mathbf{R})\right\rangle $ defined by%
\begin{equation}
\left\langle \mathcal{S}(\mathbf{R}))\right\rangle =\frac{1}{Z}\int d\mathbf{%
\Psi }\int d\mathbf{p_{\mathbf{\Psi }}}\exp (-\beta H[\mathbf{\Psi },\mathbf{%
p_{\mathbf{\Psi }}}])\mathcal{S}(\mathbf{R}).  \label{S(R)definition}
\end{equation}%
Following our discussion in previous Section, we approximate it as

\begin{equation}
\left\langle \mathcal{S}(\mathbf{R})\right\rangle \cong \frac{1}{Z_{R}}\int d%
\mathbf{\Psi }_{L}\int d\mathbf{p_{\mathbf{\Psi }_{L}}}\exp (-\beta H_{L}[%
\mathbf{\Psi },\mathbf{p_{\mathbf{\Psi }_{L}}}])\mathcal{S}(\mathbf{R})
\label{S(R)approximated}
\end{equation}%
It can be easily evaluated, following the methods of Appendix \ref%
{EvaluationofG}. \ The result is
\begin{equation}
\left\langle \mathcal{S}(\mathbf{R})\right\rangle =16\sqrt{\frac{2\left\vert
\mathbf{T}_{L}^{\phi }\right\vert }{3\left\vert \mathbf{T}_{R}^{\phi
}\right\vert }}\frac{e^{-\frac{4L^{2}\pi ^{2}\beta \kappa }{3L^{3}+8\pi
^{2}\beta \eta ^{2}\kappa }}\pi ^{3}\beta ^{2}\kappa ^{2}}{\sqrt{L}\sqrt{%
L^{3}-4\pi ^{2}\beta \eta ^{2}\kappa }\sqrt{L^{3}+8\pi ^{2}\beta \eta
^{2}\kappa }\sqrt{3L^{3}+8\pi ^{2}\beta \eta ^{2}\kappa }}.  \label{<S(R)>}
\end{equation}

\subsection{The sink-sink correlation function}

\label{sinkcorrelationfunction} The essential step in finding the average
time of loop formation is to calculate the sink-sink correlation function,
Eq. (\ref{sinkcor}). The sink-sink correlation function can be written in
terms of \ $\mathbf{\Psi }$:
\begin{equation}
\begin{array}{c}
\mathcal{D}(t)=\int d\mathbf{\Psi }\int d\mathbf{\Psi }^{\prime }\mathcal{S}(%
\mathbf{R}^{\prime })G_{0}(\mathbf{\Psi }^{\prime },\mathbf{\Psi };t)%
\mathcal{S}(\mathbf{R})P_{eq}(\mathbf{\Psi }).%
\end{array}
\label{D(t)}
\end{equation}%
$G_{0}(\mathbf{\Psi }^{\prime },\mathbf{\Psi };t)$ is the propagator
expressed in terms of $\mathbf{\Psi }$ and obeys the condition $G_{0}(%
\mathbf{\Psi }^{\prime },\mathbf{\Psi };t)\mathcal{S}(\mathbf{R})\rightarrow
\delta (\mathbf{\Psi ^{\prime }}-\mathbf{\Psi }))$ as $t\rightarrow 0$. The $%
P_{eq}(\mathbf{\Psi })$ in the above is given by
\begin{equation}
P_{eq}(\mathbf{\Psi })=\frac{1}{Z}\int d\mathbf{\Psi }\int d\mathbf{p_{%
\mathbf{\Psi }}}\exp (-\beta H[\mathbf{\Psi },\mathbf{p_{\mathbf{\Psi }}}]).
\label{PeqintermsofPsi}
\end{equation}%
In the spirit of our previous discussions, we approximate $P_{eq}(\mathbf{%
\Psi })$ near the loop configuration as%
\begin{equation*}
P_{eq}(\mathbf{\Psi })\cong \frac{1}{Z_{R}}\int d\mathbf{p_{\mathbf{\Psi }%
_{L}}}\exp (-\beta H_{L}[\mathbf{\Psi },\mathbf{p_{\mathbf{\Psi }_{L}}}]).
\end{equation*}%
The sink-sink correlation function may be written as
\begin{equation}
\mathcal{D}(t)=\left\langle \mathcal{S}(\mathbf{R})\right\rangle \mathcal{C}%
(t),  \label{D(t)intermsofS(R)andC(t)}
\end{equation}%
where
\begin{equation}
\mathcal{C}(t)=\frac{1}{\left\langle S(\mathbf{R})\right\rangle }\int d%
\mathbf{\Psi }\int d\mathbf{\Psi }^{\prime }\mathcal{S}(\mathbf{R}^{\prime
})G_{0}(\mathbf{\Psi }^{\prime },\mathbf{\Psi };t)\mathcal{S}(\mathbf{R}%
)P_{eq}(\mathbf{\Psi }).  \label{C(t)Definition}
\end{equation}%
$\mathcal{C}(t)$ \ can now be approximated as
\begin{equation}
\mathcal{C}(t)\cong \mathcal{C}_{a}(t)=\frac{\int d\mathbf{\Psi }_{L}\int d%
\mathbf{\Psi }_{L}^{\prime }S(\mathbf{R}^{\prime })G_{0}(\mathbf{\Psi }%
_{L}^{\prime },\mathbf{\Psi }_{L};t)S(\mathbf{R})\exp \left( -\beta \frac{%
\kappa }{2L}\mathbf{\Psi }_{L}^{\dag }\mathcal{V}_{L}\mathbf{\Psi }%
_{L}\right) }{\int d\mathbf{\Psi }_{L}\mathcal{S}(\mathbf{R})\exp \left(
-\beta \frac{\kappa }{2L}\mathbf{\Psi }_{L}^{\dag }\mathcal{V}_{L}\mathbf{%
\Psi }_{L}\right) }.  \label{C(t)Approximation}
\end{equation}%
Note that we use the subscript $``a"$ to denote the approximate value of $%
\mathcal{C}(t)$. The above integral may be re-expressed in terms of the
normal modes $\mathbf{Y}$ as
\begin{equation*}
\mathcal{C}_{a}(t)\mathbf{=}\frac{\int d\mathbf{Y}^{\prime }\int d\mathbf{Y}%
\mathcal{S}(\mathbf{R}^{\prime })G_{0}(\mathbf{Y}^{\prime },\mathbf{Y};t)%
\mathcal{S}(\mathbf{R})\exp \left( -\beta \frac{\kappa }{2L}\mathbf{Y}^{\dag
}\mathcal{K}\mathbf{Y}\right) }{\int d\mathbf{Y}\mathcal{S}(\mathbf{R})\exp
\left( -\beta \frac{\kappa }{2L}\mathbf{Y}^{\dag }\mathcal{K}\mathbf{Y}%
\right) }.
\end{equation*}%
$G_{0}(\mathbf{Y}^{\prime },\mathbf{Y};t)$ is the propagator expressed in \
terms of the normal modes $\mathbf{Y}$. \ The Jacobians associated with the
transformation in the numerator and denominator cancel out (Note that $\int d%
\mathbf{Y}^{\prime }G_{0}(\mathbf{Y}^{\prime },\mathbf{Y};t)=1$). \ With the
above form the sink function, $\mathcal{C}_{a}(t)$ can be evaluated to
obtain (see appendix \ref{EvluationofS(R)andC(t)})
\begin{equation}
\mathcal{C}_{a}(t)=\mathcal{C}_{x}(t)\mathcal{C}_{y}(t)\mathcal{C}_{z}(t)
\label{CaintermsofCxCyCz}
\end{equation}%
with
\begin{equation}
\mathcal{C}_{x}(t)=\frac{e^{\frac{1}{2}L^{2}\beta \kappa \left( \frac{1}{%
S_{x}(0)L^{3}+\beta \eta ^{2}\kappa }-\frac{2}{\left( S_{x}(0)+S_{x}(%
\overline{t})\right) L^{3}+\beta \eta ^{2}\kappa }\right) }\sqrt{\beta }%
\sqrt{\kappa }\sqrt{S_{x}(0)L^{3}+\beta \eta ^{2}\kappa }}{\sqrt{2\pi }\sqrt{%
\left( S_{x}(0)L^{3}+\beta \eta ^{2}\kappa \right) {}^{2}-L^{6}S_{x}(%
\overline{t}){}^{2}}},  \label{Cxt}
\end{equation}%
\begin{equation}
\mathcal{C}_{y}(t)=\frac{\sqrt{\beta }\sqrt{\kappa }\sqrt{%
S_{y}(0)L^{3}+\beta \eta ^{2}\kappa }}{2\sqrt{2}\pi ^{3/2}\sqrt{\left(
S_{y}(0)L^{3}+\beta \eta ^{2}\kappa \right) {}^{2}-L^{6}S_{y}(\overline{t}%
){}^{2}}}  \label{Cyt}
\end{equation}%
and%
\begin{equation}
\mathcal{C}_{z}(t)=\frac{\sqrt{\beta }\sqrt{\kappa }\sqrt{%
S_{z}(0)L^{3}+\beta \eta ^{2}\kappa }}{\sqrt{2}\pi ^{5/2}\sqrt{\left(
S_{z}(0)L^{3}+\beta \eta ^{2}\kappa \right) {}^{2}-L^{6}S_{z}(\overline{t}%
){}^{2}}}.  \label{Czt}
\end{equation}%
On using these,
\begin{equation}
\mathcal{C}_{a}(t)=\sqrt{\frac{3\left\vert \mathbf{T}_{R}^{\phi }\right\vert
}{2\left\vert \mathbf{T}_{L}^{\phi }\right\vert \kappa \beta \pi ^{15}}}e^{%
\frac{1}{2}L^{2}\beta \kappa \left( \frac{1}{S_{x}(0)L^{3}+\beta \eta
^{2}\kappa }-\frac{2}{\left( S_{x}(0)+S_{x}(\overline{t})\right) L^{3}+\beta
\eta ^{2}\kappa }+\frac{8\pi ^{2}}{3L^{3}+8\pi ^{2}\beta \eta ^{2}\kappa }%
\right) }\frac{\sqrt{L}}{128}
\end{equation}%
\begin{equation*}
\frac{\sqrt{\left( L^{3}-2\pi ^{2}\beta \eta ^{2}\kappa \right) \left(
L^{3}+8\pi ^{2}\beta \eta ^{2}\kappa \right) \left( 3L^{3}+8\pi ^{2}\beta
\eta ^{2}\kappa \right) }\sqrt{S_{x}(0)L^{3}+\beta \eta ^{2}\kappa }\sqrt{%
S_{y}(0)L^{3}+\beta \eta ^{2}\kappa }\sqrt{2S_{z}(0)L^{3}+2\beta \eta
^{2}\kappa }}{\sqrt{\left( S_{x}(0)L^{3}+\beta \eta ^{2}\kappa \right)
{}^{2}-L^{6}S_{x}(\overline{t}){}^{2}}\sqrt{\left( S_{y}(0)L^{3}+\beta \eta
^{2}\kappa \right) {}^{2}-L^{6}S_{y}(\overline{t}){}^{2}}\sqrt{\left(
S_{z}(0)L^{3}+\beta \eta ^{2}\kappa \right) {}^{2}-L^{6}S_{z}(t){}^{2}}}
\end{equation*}%
In the above
\begin{equation}
\begin{array}{c}
S_{x}(\overline{t})=\sum\limits_{n=1}^{\mathcal{N}}\frac{f_{n}^{2}}{k_{nx}}%
e^{-\overline{t}k_{nx}}{\mathrm{;}}S_{y}(\overline{t})=\sum\limits_{n=2}^{%
\mathcal{N}}\frac{g_{n}^{2}}{k_{ny}}e^{-\overline{t}k_{ny}}{\mathrm{;}}S_{z}(%
\overline{t})=\sum\limits_{n\neq 2}^{\mathcal{N}}\frac{h_{n}^{2}}{k_{nz}}e^{-%
\overline{t}k_{nz}}.%
\end{array}
\label{Sx,Sy,SzDefinitions}
\end{equation}%
with $\overline{t}=t/\tau _{0}$. $S_{x}(0)$, $S_{y}(0)$ and $S_{z}(0)$ can
be evaluated exactly (see Appendix \ref{Sx(0)Sy(0)Sz(0)}). Using their
values given by equations (\ref{sx0}), (\ref{sy0}) and (\ref{sz0}). \
Defining
\begin{equation}
f(\overline{t})=S_{x}(\overline{t})/S_{x}(0),g(\overline{t})=S_{y}(\overline{%
t})/S_{y}(0);\mbox{and }h(\overline{t})=S_{z}(\overline{t})/S_{z}(0).
\label{fx,fy,fzDefinitions}
\end{equation}%
we get$\ $
\begin{eqnarray}
C_{a}(t) &=&\left( \frac{2\beta \kappa }{\pi }\right) ^{3/2}e^{\frac{1}{2}%
L^{2}\beta \kappa \left( \frac{8\pi ^{2}}{3\ L^{3}+8\pi ^{2}\beta \eta
^{2}\kappa }-\frac{16\pi ^{2}}{3f(\overline{t})L^{3}+3L^{3}+8\ \pi ^{2}\beta
\eta ^{2}\kappa }\right) }  \label{Ca(t)} \\
&&\frac{\sqrt{L^{3}-4\ \pi ^{2}\beta \eta ^{2}\kappa }\sqrt{L^{3}+8\pi
^{2}\beta \eta ^{2}\ \kappa }\sqrt{3L^{3}+8\pi ^{2}\beta \eta ^{2}\kappa }}{%
\ \sqrt{\left( 3L^{3}+8\pi ^{2}\beta \eta ^{2}\kappa \right) ^{2}-9L^{6}f(%
\overline{t})^{2}}\ \sqrt{\left( L^{3}+8\pi ^{2}\beta \eta ^{2}\kappa
\right) ^{2}-L^{6}g(\overline{t})^{2}}\ \sqrt{L^{6}\ h(\overline{t}%
)^{2}-\left( L^{3}-4\pi ^{2}\beta \eta ^{2}\kappa \right) ^{2}}}  \notag
\end{eqnarray}%
The values of $f_{n}/k_{xn}^{2},g_{n}/k_{yn}^{2}$ and $h_{n}/k_{zn}^{2}$ are
given in Table \ref{table1}. \ Use of equations (\ref{<S(R)>}) and (\ref%
{Ca(t)}) in Eq. (\ref{D(t)intermsofS(R)andC(t)}) leads to an approximation
for $\mathcal{D}(t)$ which we denote by $\mathcal{D}_{a}(t)$.
\begin{equation}
\mathcal{D}_{a}(t)=\left\langle S(\mathbf{R})\right\rangle \mathcal{C}%
_{a}(t).  \label{Da(t)intermsofCa(t)}
\end{equation}%
Since the eigenvalue $k_{2z}$ is negative (see Table \ref{table1}), the term
$g(t)$ has a term that diverges exponentially as $t\rightarrow \infty $,
making $\mathcal{C}_{z}(\infty )=0$. \ Hence $\mathcal{D}_{a}(\infty )$ is
zero. This is due to the instability along $R_{z}$, which causes the
correlation function to vanish at long times. Hence, $\mathcal{D}_{a}(t)$
given by Eq. (\ref{Da(t)intermsofCa(t)}) is a good approximation to $%
\mathcal{D}(t)$ at short times. \ For long times $\mathcal{D}(t)$ will
approach $v_{eq}^{2}$. \ \ Hence $\mathcal{D}(t)\cong \mathcal{D}_{a}(t)$ is
a valid approximation only at short times. For for long times, $\mathcal{D}%
(t)$ should be equal to $v_{eq}^{2}$. Hence, it follows that the actual
correlation function may be approximated as $\mathcal{D}(t)\cong \mathcal{D}%
_{a}(t)+v_{eq}^{2}$. Using this in Eq. (\ref{sinktau}) one gets
\begin{equation}
\tau =\frac{1}{v_{eq}^{2}}\int\limits_{0}^{\infty }\mathcal{D}_{a}(t)dt.
\label{tauclose}
\end{equation}%
We note that $\beta \kappa =l_{p}$ is the persistence length of the polymer
and that $\frac{\rho (\beta \kappa )^{4}\gamma }{\kappa }$ has dimensions of
time, and use these as units for length and time. \ Then, the expression for
$\tau $ becomes
\begin{equation}
\tau =\sqrt{\frac{3\left\vert \mathbf{T}_{R}^{\phi }\right\vert }{%
64\left\vert \mathbf{T}_{L}^{\phi }\right\vert \pi ^{9}}}L^{9/2}W(L)
\label{loopingtime}
\end{equation}%
with%
\begin{eqnarray}
W(L) &=&\int_{0}^{\infty }dte^{\frac{24L^{5}\pi ^{2}\ f(t)}{\left(
3L^{3}+8\pi ^{2}\eta ^{2}\right) \left( 3f(t)L^{3}+3L^{3}+8\pi ^{2}\eta
^{2}\right) }}\left( 1-\frac{f(t)^{2}}{\left( 1+8\pi ^{2}\eta
^{2}/3L^{3}\right) ^{2}}\right) ^{-1/2}  \label{W(L)a} \\
&&\left( 1-\frac{\ g(t)^{2}}{\left( 1+8\pi ^{2}\eta ^{2}/L^{3}\right) ^{2}}%
\right) ^{-1/2}\left( \frac{h(t)^{2}}{\left( 1-4\pi ^{2}\eta
^{2}/L^{3}\right) ^{2}}-1\right) ^{-1/2}.  \notag
\end{eqnarray}%
The functions $f(t),$ $g(t)$ and $h(t)$, on evaluation are found to be given
by%
\begin{equation}
f(t)=0.995873e^{-298.541t}+0.00408587e^{-7126.56t}.  \label{fx(tbar)}
\end{equation}%
\begin{eqnarray}
g(t) &=&0.921041e^{-1413.12t}+0.0625692e^{-24548.8t}+0.0105256e^{-134918.t}
\notag \\
&&+0.00312111e^{-436308.t}+0.00124791e^{-1.06959\times 10^{6}t}.
\label{fy(tbar)}
\end{eqnarray}%
\begin{eqnarray}
h(t) &=&1.02629e^{1241.14t}-0.0197683e^{-28705.1t}-0.00396455e^{-152588.t}
\notag \\
&&-0.00130213e^{-474114.t}.  \label{fz(tbar)}
\end{eqnarray}%
Terms that make no significant contribution to $f(t)$, $g(t)$ and $h(t)$, as
their exponents are large and the coefficients small have been neglected in
the above. Using Eq. (\ref{loopingtime}), we have calculated the average
time of loop formation $\tau $ as a function of the length $L$, and the
results are given in figures \ref{loopingtime1} and \ref{loopingtime2}. The
results are dependent on the sink width $\eta $. \ The full lines in these
figures give exact values of $\tau $ for various values of sink size $\eta .$
The values of $\tau $ for each value of $\eta $ were fitted with an equation
of the form $A\exp (E_{a}/L)L^{n}$ and the values of $E_{a}$ and $n$, as
well as the value of the length at which $\tau $ is a minimum ($L_{m}$) is
given in Table \ref{fittedparameters}. \ The curves that result from the
fitting are shown as dotted lines in the figures. \ It is seen that the
functional form $\tau =A\exp (E_{a}/L)L^{n}$ reproduces the data well, with the exponent in the prefactor $%
n$ being dependent on $\eta $ and varying from $4.5$ to $6$. \ For small $%
\eta ,$  $n$ is close to $6$. \ Further, the loop formation time $\tau $
becomes longer and longer as $\eta $ is made smaller. In fact, as may be
seen by from Eq. (\ref{loopingtime}), the time diverges like $1/\eta $ as $%
\eta \rightarrow 0$ (see Eq. (\ref{tauetaapproaches0})). This is not
surprising, because as $\eta \rightarrow 0$, the diffusional search for loop
formation is for a smaller and smaller volume in space. In fact, on looking
at Eq. (\ref{loopingtime}) and remembering that $\eta $ actually serves the
purpose of a small length cut off, one would have expected the dependence to
be approximately
\begin{equation}
\tau =A\exp (E_{a}/L)L^{9/2}.  \label{simpleform}
\end{equation}%
However, there are two reasons that lead to the observed dependence on $\eta
$. (1) The exponential term in Eq. (\ref{W(L)a}) is dependent on $t$ and
makes the $L$ dependence change from the simple form of Eq. (\ref{simpleform}%
) and (2) the $L$ dependence of the terms inside the square roots in Eq. (%
\ref{W(L)a}). \ From the functional forms of $f(t),$ $g(t)$ and $h(t),$ it
is clear that due to the presence of $h(t),$ the integrand decreases rather
rapidly. For $t\sim 1/\omega _{h},$where $\omega _{h}=1241.4$ (see Eq. \ref%
{fx(tbar)}) $\left( \frac{h(t)^{2}}{\left( 1-4\pi ^{2}\eta ^{2}/L^{3}\right)
^{2}}-1\right) ^{-1/2}\cong h(t)^{-1}=$ $(1/1.02629)e^{-\omega _{h}t}$ . \ \
On this time scale ($t\sim 1/\omega _{h}$), one can approximate $\left( 1-%
\frac{\ g(t)^{2}}{\left( 1+8\pi ^{2}\eta ^{2}/L^{3}\right) ^{2}}\right)
^{-1/2}\cong 1$ and $\left( 1-\frac{f(t)^{2}}{\left( 1+8\pi ^{2}\eta
^{2}/3L^{3}\right) ^{2}}\right) ^{-1/2}$ $\cong \left( 3L^{3}/16\pi ^{2}\eta
^{2}\right) ^{1/2}.$ \ Hence for small values of $\eta $, the integral may
be evaluated approximately to get\
\begin{equation}
\tau _{close}(L)=\frac{3}{32\omega _{h}\eta }\sqrt{\frac{\left\vert \mathbf{T%
}_{R}^{\phi }\right\vert }{\left\vert \mathbf{T}_{L}^{\phi }\right\vert \pi
^{11}}}L^{6}e^{\frac{4\pi ^{2}\ }{3L}}.  \label{tauetaapproaches0}
\end{equation}%
Thus for small $\eta $ the loop formation time behaves like $\sim L^{6}e^{%
\frac{4\pi ^{2}\ }{3L}}$ as seen in the Table \ref{fittedparameters}. For
not so small values of $\eta ,$ one expects $\left( 1-\frac{f(t)^{2}}{\left(
1+8\pi ^{2}\eta ^{2}/3L^{3}\right) ^{2}}\right) ^{-1/2}$ $\cong 1$ and $%
\left( 1-\frac{\ g(t)^{2}}{\left( 1+8\pi ^{2}\eta ^{2}/L^{3}\right) ^{2}}%
\right) ^{-1/2}\cong 1$. \ This leads to  $n\sim 9/2$. \

Physically, the above results are easy to understand. \ The rate of the loop
formation may be written as $\simeq \mathcal{P}(R_{x}\lesssim \eta
,R_{y}\lesssim \eta ,R_{z}\lesssim \eta )\times $ frequency factor $\times
Z_{L}/Z_{R}$. \ \ In this, $\mathcal{P}(\mathbf{R})$\ is the probablity
distribution function for the end-to-end vector. \ For a semi-flexible
chain, the frequency factor $\sim L^{-4}.$ \ In the limit $\eta \rightarrow
0,$ $\mathcal{P}(R_{x}\lesssim \eta ,R_{y}\lesssim \eta ,R_{z}\lesssim \eta
)\sim $ $(\beta \kappa /L)^{3/2}$ as seen earlier and $Z_{L}/Z_{R}\sim
(\beta \kappa /L)^{1/2}$. \ Therefore, the preexponential factor of the rate has $(\beta \kappa
)^{2}/L^{6} $ dependence. \ On the other hand, as one increases the value of
$\eta $, for sufficiently large $\eta $, $\mathcal{P}(R_{x}\lesssim \eta
,R_{y}\lesssim \eta ,R_{z}\lesssim \eta )=1$, leading to rate of the form $%
(\beta \kappa )^{1/2}/L^{9/2}$\cite{referee}.

\subsection{\protect\bigskip Numerical Results}

\label{numericalresults}
\begin{figure}[tbp]
\includegraphics{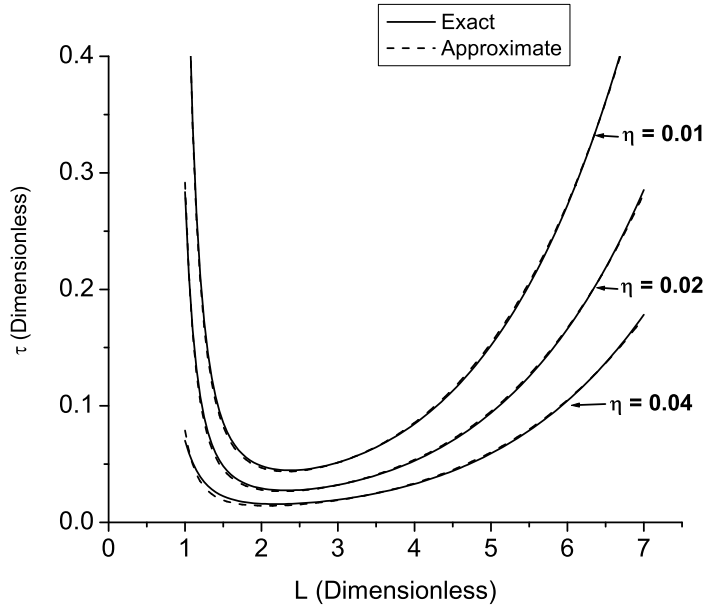}
\caption{ The time of loop formation, $\protect\tau $ as a function of $L$,
for different values of the width $\protect\eta $. Units are chosen such
that both $\protect\tau $ and $L$ dimensionless. The full curves are the
computed results. They are well represented by the functional form $%
AL^{n}\exp (E_{a}/L)$ as may be seen from the figure, where we have
represented them by dotted lines. The parameters that result from fitting
are given in Table \protect\ref{fittedparameters}.}
\label{loopingtime1}
\end{figure}

\begin{figure}[tbp]
\includegraphics[scale = 1]{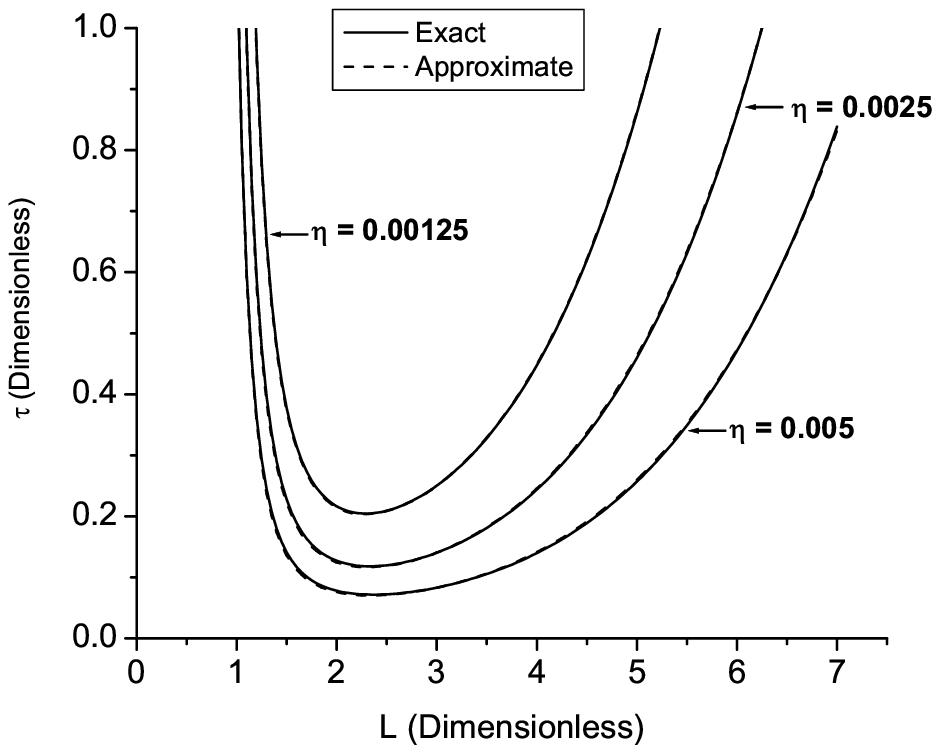}
\caption{ The closing time $\protect\tau$ as a function of $L$, for
different values of the width $\protect\eta$. Units are adopted such that
all these are dimensionless. The full curves are the computed results. They
are well represented by the functional form $A L^n \exp(E_a/L) $ as may be
seen from the figure, where we have represented them by dotted lines. The
parameters that result from fitting are given in Table \protect\ref%
{fittedparameters}.}
\label{loopingtime2}
\end{figure}

\begin{table}[tbp]
\begin{tabular}{|c|c|c|c|c|}
\hline
$\eta$ &$A \times 10^6$& $E_a$ & $n$ & $L_m$ \\ \hline
0.04 & 2.800 &10.247 & 4.920 & 2.19 \\
0.02 &1.823& 11.982 & 5.261 & 2.32 \\
0.01 & 1.627&12.890 & 5.51 & 2.37 \\
0.005 & 1.740&13.421 & 5.74 & 2.37 \\
0.0025 &2.268& 13.64 & 5.90 & 2.33 \\
0.00125 & 3.670&13.65 & 5.99 & 2.28 \\ \hline
\end{tabular}%
\caption{Fitted Parameters: Values of $A$, $E_{a}$ and $n$ and $L_{m}$ (the
dimensionless length at which $\protect\tau $ is a minimum).}
\label{fittedparameters}
\end{table}

We now consider loop formation of double stranded DNA, which has a
persistence length of $50$ nm.
The sink function, defined by Eq. (\ref{S(R)}) and (\ref{SiRi}) has a width
equal to $\eta $ which may be taken as $\ 2-3$ \AA . \ \ Then the
dimensionless $\eta $ would have a value of roughly $1/200$ and this would
correspond to the lowest curve in Fig. \ref{loopingtime2}, with $n=5.9$. \
On the other hand for a more flexible chain, with persistence length equal
to $2.5$ nm, and with a value of $\eta $ equal to $1$ \AA , one would have
dimensionless $\eta =1/25,$ and this would correspond to the lowest curve in
Fig. \ref{loopingtime1} with $n=4.9$. \ \ The value of $L_{m}$ at which the
minmum time is required for loop formation, does not depend strongly on the
value of $\eta $. \ Thus it is found to be in the range $2.2$ to $2.4$ times
the persistence length of the chain (see Table \ref{fittedparameters}). \

The dynamics of loop formation in semiflexible polymers was analyzed by Dua
and Cherayil, who found $\tau \sim L^{n},$ with $n$ $\sim 2.2-2.4$, with $n$
approaching $2$ in the flexible limit. \ This is obviously valid in the
longer chain limit. \ On the other hand, Jun et al \cite%
{junPC2003,junEPL2003} have studied the region where the length of the chain
is a few times the persistence length. \ The assumed the two ends of the
chain to execute random walk with a constant diffusion coefficient, and
found that there is a length ($L_{m}$) at which $\tau $ is a minimum. \ \
Their analysis used accurate results for $G(\mathbf{0},L)$ and lead to
somewhat larger value for $L_{m}$ ($3-4$).\ On the other hand, we have
studied the dynamics in detail, using a multimode approach. \ We get
expressions for $G(\mathbf{0},L)$ and $\tau $ which have extrema at lower
values of $L$, this being a result of the use of approximate expression for
the bending energy.

\section{Summary and Conclusions}

\label{secsix} In this work, we have presented a detailed multi-dimensional
analysis of the loop formation dynamics of semiflexible chains. The reverse
process, the opening of the loop was studied in a previous work \cite%
{klssantoPRE2006}, where we developed an approximate model for a
semiflexible chain in the rod limit. In this model, the conformations of the
polymer are mapped onto the paths of a random walker on the surface of a
unit sphere. The bending energy of the chain was expanded about a minimum
energy path. This model was shown to be a good approximation for the polymer
in the rod limit and provided an easy way to describe the dynamics. Use of
this model led to opening rates of a semiflexible polymer loop formed by a
weak bond between the ends. in Ref. \cite{klssantoPRE2006}, we calculated
the opening rates for a Morse type interaction between the ends of the
polymer as a function of the contour length of the chain. In this paper, we
analyzed the loop formation dynamics using this model and thus, presented a
rather complete theory of dynamics of formation of semiflexible polymer
loops.

The dynamics was described using the formalism by Wilemski and Fixman, which
describe the intra-chain reactions of polymers as a diffusion-controlled
reaction. In this formalism, the reaction process is described using a sink
function. For an arbitrary sink function, exact results are not available
and hence, WF introduced an approximation called ``closure approximation".
In this procedure, the closing time can be expressed in terms of a sink-sink
correlation function. To calculate this sink-sink correlation function and
thereby the closing time, one needs to know the Green's function of the
chain and the equilibrium distribution. We calculated the Green's function
of the chain through a normal mode analysis near the loop geometry. This
normal mode analysis could be performed independently of the rigidity ($%
\kappa$) and contour length ($L$) of the polymer, leading to a set of
eigenvalues that are universal. An approximate equilibrium distribution for
the polymer near the ring configuration was given. As the sink function
vanishes for large values of the end-to-end distance $R$, sink-sink
correlation function has contributions mostly from the dynamics of the
polymer near the ring configurations. We calculated this approximate
sink-sink correlation function for a Gaussian sink through a transformation
of variables into normal coordinates.

We then obtained loop formation time (in dimensionless units), $\tau $ for
different contour lengths of the chain. We found that $\tau \sim L^{9/2}W(L)$%
, where $W(L)$ is an integral that could be performed numerically. Numerical
calculations lead to the result that $\tau =AL^{n}\exp (E_{a}/L)$, with $n$
varying between $9/2$ and $6$.  $\tau $ was found to have a minimum at $%
L_{min}=2.2$ to $2.4$ which is to be compared with the values $3.4$ obtained
by Jun et. al. \cite{junEPL2003} by a simple one dimensional analysis and
the value  $2.85$ of Chen et. al. \cite{chen04EL407} found\ through
simulations \cite{chen04EL407}.

\begin{center}
\textbf{ACKNOWLEDGEMENTS}
\end{center}

K. P. Santo thanks Council of Scientific and Industrial Research (CSIR),
India for financial support and Cochin University of Science and Technology
(CUSAT), Kochi, India for providing computer facilities during the
preparation of the manuscript. \ The work of KLS is supported by the J.C.
Bose Fellowship of the Department of Science and Technology, Govt. of India.

\appendix

\section{The kinetic energy of the ring and the rod}

\label{AppendixKEMatrices} The kinetic energy Eq. (\ref{kepoly}) of the
polymer may be evaluated using Eq. (\ref{positionvector}). We take $\mathbf{r%
}_{cm}=0$, so that the ring is described in the center of mass frame and the
translational degrees of freedom are eliminated.

\subsection{Matrix elements for the Loop}

The kinetic energy matrix elements of the loop are given below.

\subsubsection{Odd $\protect\phi $ Modes}

For odd $\delta \phi _{n}$ modes one has
\begin{equation}
(\mathbf{T}_{L}^{\phi o})_{mn}=t_{n}\delta _{mn}-16t_{n}t_{m},
\label{TLoopmnphio}
\end{equation}%
where
\begin{equation}
t_{n}=\frac{4+n^{2}}{2(-4+n^{2})^{2}\pi ^{2}}.  \label{tn}
\end{equation}

\subsubsection{Even $\protect\phi $ Modes}

For even $\delta \phi _{n}$, one has
\begin{equation}
(\mathbf{T}_{L}^{\phi e})_{mn}=t_{n}\delta _{mn}  \label{TLoopmnphieven}
\end{equation}%
with $n,m\neq 2$. For $m,n=2$ one gets
\begin{equation}
(\mathbf{T}_{L}^{\phi e})_{22}=\frac{-3+4\pi ^{2}}{192\pi ^{2}}
\label{TLoopphi22}
\end{equation}%
and
\begin{equation}
(\mathbf{T}_{L}^{\phi e})_{2n}=-\frac{4+n^{2}}{2(-4+n^{2})^{2}\pi ^{2}}=(%
\mathbf{T}_{L}^{\phi e})_{n2}  \label{TLoopphi2m}
\end{equation}

\subsubsection{Odd $\protect\theta $ Modes}

The kinetic energy matrix corresponding to the even $\theta $ modes have the
following form. For $n,m\neq 0$
\begin{equation}
(\mathbf{T}_{L}^{\theta o})_{nm}=q_{n}\delta _{mn}  \label{TLooptheta}
\end{equation}%
where
\begin{equation}
q_{n}=\frac{1}{2n^{2}\pi ^{2}}.  \label{qn}
\end{equation}%
For the zeroth mode
\begin{equation}
(\mathbf{T}_{L}^{\theta o})_{00}=\frac{1}{12}  \label{TLooptheta00}
\end{equation}%
and
\begin{equation}
(\mathbf{T}_{L}^{\theta o})_{0n}=-\frac{1}{n^{2}\pi ^{2}}=(\mathbf{T}%
_{L}^{\theta })_{n0}.  \label{TLooptheta0n}
\end{equation}

\subsubsection{Even $\protect\theta $ Modes}

The kinetic energy matrix corresponding to the odd $\theta $ modes have the
following form
\begin{equation}
(\mathbf{T}_{L}^{\theta e})_{nm}=q_{n}\delta _{mn}-d_{n}d_{m},
\end{equation}%
where
\begin{equation}
q_{n}=\frac{1}{2n^{2}\pi ^{2}}
\end{equation}%
and
\begin{equation}
d_{n}=-\frac{2}{n^{2}\pi ^{2}}.
\end{equation}

\subsection{For the Rod}

In the case of a rod, the matrix elements are identical for $\phi $ and $%
\theta $ modes. \ They are given by
\begin{equation}
(\mathbf{T}_{R}^{\phi o})_{mn}=(\mathbf{T}_{R}^{\theta
o})_{mn}=t_{n}^{\prime }\delta _{mn}-16t_{n}^{\prime }t_{m}^{\prime },
\label{TRodmnphiodd}
\end{equation}%
where
\begin{equation}
t_{n}^{\prime }=\frac{1}{2n^{2}\pi ^{2}}.  \label{tnp}
\end{equation}%
For modes with even $n$ one has
\begin{equation}
(\mathbf{T}_{R}^{\phi e})_{mn}=(\mathbf{T}_{R}^{\theta
e})_{mn}=t_{n}^{\prime }\delta _{mn}  \label{TRodmnphieven}
\end{equation}%
with $n,m\neq 0$. In this case, the zeroth mode is couples to the other
modes. Thus one gets
\begin{equation}
(\mathbf{T}_{R}^{\phi e})_{00}=(\mathbf{T}_{R}^{\theta e})_{00}=\frac{1}{12}
\label{TRodphi00}
\end{equation}%
and
\begin{equation}
(\mathbf{T}_{R}^{\phi e})_{0n}=(\mathbf{T}_{R}^{\theta e})_{0n}=\frac{1}{%
n^{2}\pi ^{2}}=(\mathbf{T}_{R}^{\phi e})_{n0}.  \label{TRodphi20n}
\end{equation}

\section{{The evaluation of $G(\mathbf{0,}L)$}}

We now give details of the evaluation of $G(\mathbf{0,}L)$ \label%
{EvaluationofG}. We perform the integral in Eq. (\ref{G(0,L,0)}) and
substitute the value of $Z_{R}$ from Eq. (\ref{Z_R}) to get
\begin{eqnarray}
G(\mathbf{0,}L\mathbf{)} &=&\frac{8\pi ^{2}}{Z_{R}}\int d\mathbf{p}_{\mathbf{%
\Psi }_{L}}\int d\mathbf{\Psi }_{L}\exp (-\beta H_{L})\delta (\mathbf{R)}%
\delta (\delta \phi _{0})\delta (\delta \theta _{2c})\delta (\delta \theta
_{2s}).  \label{G(0,L|0)appendix1} \\
&=&\pi 2^{3-4\mathcal{N}}\left( \frac{\kappa \beta \pi }{3}\right) ^{2%
\mathcal{N}-1}\sqrt{\frac{\left\vert \mathbf{T}_{L}^{\phi }\right\vert }{%
\left\vert \mathbf{T}_{R}^{\phi }\right\vert }}G^{\phi o}G^{\phi e}G^{\theta
o}G^{\theta e},
\end{eqnarray}%
where $G^{\phi o}G^{\phi e}G^{\theta o}G^{\theta e}$ are defined and
calculated in the following. \ $G^{\phi o}$ is the contribution from the odd
$\phi $ modes to $G(\mathbf{0,}L\mathbf{)}$ and is defined by
\begin{equation}
G^{\theta o}=\prod\limits_{n,odd}\int d\delta \theta _{n}\exp \left( -\frac{%
\beta \kappa \pi ^{2}}{4L}(n^{2}-4)\delta \theta _{n}^{2}\right) \delta
(\delta \theta _{2s}).  \label{Gthetaodddefinition}
\end{equation}%
We put $\delta (\delta \theta _{2s})=(1/2\pi )\int dp\exp (ip\delta \theta
_{2s})=(1/2\pi )\int dp\exp (2pi\sum\limits_{n,odd}a_{n}\delta \theta _{n})$%
. \ With this, integrals over $\delta \theta _{n}$ with $n=3,5,..$ are
evaluated and then the one over $p$, after which one can easily evaluate the
integral over $\delta \theta _{1}.$ The result is
\begin{equation}
G^{\theta o}=\sqrt{\frac{\pi }{3}}\left( \frac{\pi \kappa \beta }{L}\right)
^{(1-\mathcal{N})/2}\frac{2}{\sqrt{\Gamma (\mathcal{N}-1/2)\Gamma (\mathcal{N%
}+3/2)}}  \label{Gthetaodd}
\end{equation}%
$\Gamma $ is the Gamma function. $\ G^{\theta e}$ is the contribution from
the even $\theta $ modes to $G(\mathbf{0,}L\mathbf{)}$ and is defined by
\begin{equation*}
G^{\theta e}=\prod\limits_{n,even}\int d\delta \theta _{n}\exp \left( -\frac{%
\beta \kappa \pi ^{2}}{4L}(n^{2}-4)\delta \theta _{n}^{2}\right) \delta
(\delta \theta _{2})\delta (L\delta \theta _{0})
\end{equation*}%
The $\delta (L\delta \theta _{0})$ comes as the $\delta (R_{z})$ part of $%
\delta (\mathbf{R})$ in Eq. (\ref{G(0,L|0)appendix1}). \ The integrals are
easy and the result is
\begin{equation}
G^{\theta e}=\sqrt{\frac{2\pi \kappa \beta }{L^{3}}}\left( \frac{\pi \kappa
\beta }{L}\right) ^{(1-\mathcal{N})/2}\frac{1}{\sqrt{\Gamma (\mathcal{N}%
-1)\Gamma (\mathcal{N}+1)}}  \label{Gthetaeven}
\end{equation}%
$G^{\phi e}$ is the contribution from the even $\phi $ modes to $G(\mathbf{0,%
}L\mathbf{)}$ and is defined by
\begin{equation}
G^{\phi e}=\prod\limits_{n,even}\int d\delta \phi _{n}\exp \left( -\frac{%
\beta \kappa \pi ^{2}}{4L}n^{2}\delta \phi _{n}^{2}\right) \delta (\delta
\phi _{0})\delta (L\delta \phi _{2}/2)  \label{GphievenDevinition}
\end{equation}%
The $\delta (\delta \phi _{0})$ comes as a result of removing the rotational
mode $\delta \phi _{0}.$ (\ref{G(0,L|0)appendix1}). \ The integrals are easy
and the result is
\begin{equation}
G^{\phi e}=\frac{2}{L}\left( \frac{\pi \kappa \beta }{L}\right) ^{(1-%
\mathcal{N}/2)}\frac{1}{\Gamma (\mathcal{N})}  \label{Gphieven}
\end{equation}%
$G^{\phi o}$ is the contribution from the odd $\phi $ modes to $G(\mathbf{0,}%
L\mathbf{)}$ and is defined by
\begin{equation*}
G^{\phi o}=\prod\limits_{n,odd}\int d\delta \phi _{n}^{\prime }\exp \left( -%
\frac{\beta \kappa \pi ^{2}}{4L}n^{2}(\delta \phi _{n}^{^{\prime
}})^{2}\right) \delta \left( L\sum_{n,odd}a_{n}\left( \delta \phi
_{n}^{\prime }+\frac{8}{n^{2}\pi }\right) \right)
\end{equation*}%
On evaluation, we get
\begin{equation}
G^{\phi o}=\frac{2}{\sqrt{3\beta \kappa L}}\left( \frac{\pi \kappa \beta }{L}%
\right) ^{(1-\mathcal{N})/2}\frac{1}{\Gamma (\mathcal{N}+1/2)}\exp (-\frac{%
4\pi ^{2}\kappa \beta }{3L})  \label{Gphiodd}
\end{equation}%
The equations (\ref{Gthetaodd}), (\ref{Gthetaeven}), (\ref{Gphiodd}) and (%
\ref{Gphieven}) above combined together with (\ref{G(0,L|0)appendix1}) \ and
with $\mathcal{N}\rightarrow \infty $ taken, gives
\begin{equation}
G(\mathbf{0,}L\mathbf{)=}\frac{16\sqrt{2}\pi ^{3}\beta ^{2}\kappa ^{2}}{%
3L^{5}}\sqrt{\frac{\left\vert \mathbf{T}_{L}^{\phi }\right\vert }{\left\vert
\mathbf{T}_{R}^{\phi }\right\vert }}\exp \left( -\frac{4\pi ^{2}\beta \kappa
}{3L}\right)   \label{G(0,L)AnswerAppendix}
\end{equation}%
The ratio $\sqrt{\frac{\left\vert \mathbf{T}_{L}^{\phi }\right\vert }{%
\left\vert \mathbf{T}_{R}^{\phi }\right\vert }}$ can be evaluated using
MATHEMATICA, taking each to be $1000\times 1000$ matrices. \ It evaluates to
\begin{equation}
\sqrt{\frac{\left\vert \mathbf{T}_{L}^{\phi }\right\vert }{\left\vert
\mathbf{T}_{R}^{\phi }\right\vert }}=6.5083.  \label{TRphiTLphiratio}
\end{equation}

\section{The evaluation of $\left\langle S(\mathbf{R})\right\rangle $ and $%
\mathcal{C}(t).$}

\label{EvluationofS(R)andC(t)} \ The evaluation of $\left\langle S(\mathbf{R}%
)\right\rangle $ and $\mathcal{C}(t)$ are similar to the evaluation of $G(%
\mathbf{0},L)$ carried out in Appendix \ref{EvaluationofG}. \ The result for
$\left\langle S(\mathbf{R})\right\rangle $ is%
\begin{equation}
\left\langle S(\mathbf{R})\right\rangle =\sqrt{\frac{2\left\vert \mathbf{T}%
_{L}^{\phi }\right\vert }{3\left\vert \mathbf{T}_{R}^{\phi }\right\vert }}%
\frac{16e^{-\frac{4L^{2}\pi ^{2}\beta \kappa }{3L^{3}+8\pi ^{2}\beta \eta
^{2}\kappa }}\pi ^{3}\beta ^{2}\kappa ^{2}}{\sqrt{L}\sqrt{L^{3}-4\pi
^{2}\beta \eta ^{2}\kappa }\sqrt{L^{3}+8\pi ^{2}\beta \eta ^{2}\kappa }\sqrt{%
3L^{3}+8\pi ^{2}\beta \eta ^{2}\kappa }}
\end{equation}%
$\mathcal{C}(t)$ may be written as a product of three terms, as already seen
in Eq. (\ref{CaintermsofCxCyCz}). \ We give expressions for them the
following. \
\begin{equation*}
\mathcal{C}_{x}(t)=\frac{\left( \prod\limits_{n=1}^{\mathcal{N}}\int
dY_{n}^{\prime }\int dY_{n}G(Y_{n}^{\prime }\mathbf{,}t|Y_{n},0\mathbf{)}%
\exp \left( -\frac{\beta \kappa }{2L}k_{nx}Y_{n}^{2}\right) \right)
S_{x}\left( L\sum\limits_{l=1}^{\mathcal{N}}f_{l}Y_{l}^{\prime }+L\right)
S_{x}\left( L\sum\limits_{m=1}^{\mathcal{N}}f_{m}Y_{m}+L)\right) }{\left(
\prod\limits_{n=1}^{\mathcal{N}}\int dY_{n}\exp \left( -\frac{\beta \kappa }{%
2L}k_{nx}Y_{n}^{2}\right) \right) S_{x}\left( L\sum\limits_{m=1}^{\mathcal{N}%
}f_{m}Y_{m}+L)\right) }.
\end{equation*}%
\ $\mathcal{C}_{y}(t)$ and $\mathcal{C}_{z}(t)$ are defined by
\begin{equation*}
\mathcal{C}_{y}(t)=\frac{\left( \prod\limits_{n=\mathcal{N}+1}^{2\mathcal{N}%
}\int dY_{n}^{\prime }\int dY_{n}G(Y_{n}^{\prime }\mathbf{,}t|Y_{n},0\mathbf{%
)}\exp \left( -\frac{\beta \kappa }{2L}k_{nx}Y_{n}^{2}\right) \right)
S_{y}\left( L\sum\limits_{l=\mathcal{N}+1}^{2\mathcal{N}}g_{l}Y_{l}^{\prime
}\right) S_{y}\left( L\sum\limits_{m=\mathcal{N}+1}^{2\mathcal{N}%
}g_{m}Y_{m}\right) }{\left( \prod\limits_{n=\mathcal{N}+1}^{2\mathcal{N}%
}\int dY_{n}\exp \left( -\frac{\beta \kappa }{2L}k_{nx}Y_{n}^{2}\right)
\right) S_{y}\left( L\sum\limits_{m=\mathcal{N}+1}^{2\mathcal{N}%
}g_{m}Y_{m}\right) }.
\end{equation*}%
\begin{equation*}
\mathcal{C}_{z}(t)=\frac{\left( \prod\limits_{n=2\mathcal{N}+1}^{3\mathcal{N}%
}\int dY_{n}^{\prime }\int dY_{n}G(Y_{n}^{\prime }\mathbf{,}t|Y_{n},0\mathbf{%
)}\exp \left( -\frac{\beta \kappa }{2L}k_{nx}Y_{n}^{2}\right) \right)
S_{z}\left( L\sum\limits_{l=2\mathcal{N}+1}^{3\mathcal{N}}h_{l}Y_{l}^{\prime
}\right) S_{z}\left( L\sum\limits_{m=2\mathcal{N}+1}^{3\mathcal{N}%
}g_{m}Y_{m}\right) }{\left( \prod\limits_{n=2\mathcal{N}+1}^{3\mathcal{N}%
}\int dY_{n}\exp \left( -\frac{\beta \kappa }{2L}k_{nx}Y_{n}^{2}\right)
\right) S_{z}\left( L\sum\limits_{m=2\mathcal{N}+1}^{3\mathcal{N}%
}h_{m}Y_{m}\right) }.
\end{equation*}%
On performing the integrations, one gets $\mathcal{C}_{x}(t),\mathcal{C}%
_{y}(t)$ and $\mathcal{C}_{z}(t)$ given in equations (\ref{Cxt}) to (\ref%
{Czt}).

\section{The evaluation of $S_{x}(0),$ $S_{y}(0)$ and $S_{z}(0)$}

\label{Sx(0)Sy(0)Sz(0)} The value of $S_{x}(0)$ can be found as follows.
Defining the vectors $\mathbf{f}=(f_{1},f_{2},....f_{\mathcal{N}})$ and $%
\mathbf{a}=(a_{1},a_{3},....a_{{2\mathcal{N}}-1})$, $S_{x}(0)$ is given by
\begin{equation}
S_{x}(0)=\sum\limits_{n=1}^{\mathcal{N}}\frac{f_{n}^{2}}{k_{nx}}
\label{sx00}
\end{equation}%
Using Equations (\ref{DiagonalizeVeff}) and (\ref{Definitionoffn}), we get
\begin{equation}
\begin{array}{c}
S_{x}(0)=\mathbf{a.(T}_{L}^{\phi o})^{-1/2}\mathbf{U}^{\phi o}\left( \mathbf{%
U}^{\phi o\dagger }\mathbf{(T}_{L}^{\phi o})^{-1/2}\mathbf{V}^{\phi o}%
\mathbf{(T}_{L}^{\phi o})^{-1/2}\mathbf{U}^{\phi o}\right) ^{-1}\mathbf{U}%
^{\phi o\dagger }\mathbf{(T}_{L}^{\phi o})^{-1/2}.\mathbf{a}^{\dagger } \\
=\mathbf{a.}\left( \mathbf{V}_{L}^{\phi o}\right) ^{-1}.\mathbf{a}^{\dagger }
\\
=\sum\limits_{n=odd}^{\infty }\frac{2a_{n}^{2}}{n^{2}\pi ^{2}}=\frac{3}{8\pi
^{2}}.%
\end{array}
\label{sx0}
\end{equation}%
Similarly, using Eq. (\ref{DiagonalizeVeff}) and (\ref{Definitionofgn})
gives
\begin{eqnarray}
S_{y}(0) &=&\sum\limits_{n}\frac{g_{n}^{2}}{k_{ny}}=  \label{sy0} \\
&&\frac{1}{4}\left[ \mathbf{(T}_{L}^{\phi e})^{-1/2}\mathbf{U}^{\phi
e}\left( \mathbf{U}^{\phi e\dagger }\mathbf{(T}_{L}^{\phi e})^{-1/2}\mathbf{V%
}_{L}^{\phi e}\mathbf{(T}_{L}^{\phi e})^{-1/2}\mathbf{U}^{\phi e}\right)
^{-1}\mathbf{U}^{\phi e\dagger }\mathbf{(T}_{L}^{\phi o})^{-1/2}\right] _{22}
\\
&=&\frac{1}{4}(\mathbf{V}_{L}^{\phi e})_{22}^{-1}=\frac{1}{8\pi ^{2}},
\end{eqnarray}%
and in a similar fashion
\begin{equation}
S_{z}(0)=\frac{1}{4}(\mathbf{V}_{1}^{\theta e})_{00}^{-1}=\frac{-1}{4\pi ^{2}%
}.  \label{sz0}
\end{equation}

\bibliographystyle{apsrev}
\bibliography{closenov7}

\end{document}